\documentclass[fleqn,usenatbib,useAMS]{mnras}
\usepackage{graphicx}	
\usepackage{amsmath}	
\usepackage{amssymb}	
\usepackage{multicol}        
\usepackage{bm}		
\usepackage{pdflscape}	
\DeclareGraphicsExtensions{.pdf,.png,.jpg,.jpeg,.eps}
\graphicspath{{./Figures/}}
\usepackage{acronym}
\usepackage{subcaption}
\usepackage{xcolor}
\usepackage[most]{tcolorbox}
\usepackage{hyperref}

\usepackage[T1]{fontenc}
\usepackage{ae,aecompl}

\usepackage{txfonts}
\usepackage{float}
\usepackage{caption}
\usepackage{lineno}

\title[Millisecond magnetars: Spin evolution and GWs]{Gravitational waves from mountains in newly born millisecond magnetars}

\author[Ankan Sur]{Ankan Sur\thanks{Contact e-mail: \href{mailto:ankansur@camk.edu.pl}{ankansur@camk.edu.pl}}and Brynmor Haskell
	\\
	\\
	Nicolaus Copernicus Astronomical Center, Polish Academy of Sciences, Bartycka 18, 00-716 Warsaw, Poland
    }
\date{Last updated \today; in original form \today}

\begin{document}
	
\newcommand{\gcc}{\ensuremath{\mbox{gm cm}^{-3}}}
\newcommand{\ergs}{\ensuremath{\mbox{erg s}^{-1}}}
\newcommand{\rstar}{\ensuremath{\mbox{$R_{\star}$}}}
\newcommand{\alfven}{\ensuremath{\mbox{Alfv\'{e}n}}}
\newcommand{\msun}{\ensuremath{\mbox{M_{\odot}}}}
	\maketitle
	
\begin{abstract}
	In this paper we study the spin-evolution and gravitational-wave luminosity of a newly born millisecond magnetar, formed either after the collapse of a massive star or after the merger of two neutron stars.
In both cases we consider the effect of fallback accretion, and consider the evolution of the system due to the different torques acting on the star, namely the spin up torque due to accretion and spin-down torques due to magnetic dipole radiation, neutrino emission, and gravitational wave emission linked to the formation of a `mountain' on the accretion poles.
Initially the spin period is mostly affected by the dipole radiation, but at later times accretion spin the star up rapidly.
We find that a magnetar formed after the collapse of a massive star can accrete up to  1 $M_{\odot}$, and survive on the order of 50 s before collapsing to a black hole. The gravitational wave strain, for an object  located at $1$ Mpc, is $h_c \sim 10^{-23}$ at kHz frequencies, making this a potential target for next generation ground based detectors.
A magnetar formed after a binary neutron star merger, on the other hand, accretes at the most 0.2 $M_{\odot}$, and emits gravitational waves with a lower maximum strain of the order of $h_c \sim 10^{-24}$, but also survives for much longer times, and may possibly be associated with the X-ray plateau observed in the light curve of a number of short gamma-ray burst.


\end{abstract}


\section{Introduction\label{sec:introduction}}
Millisecond magnetars are suggested to be rapidly rotating neutron stars (NSs) with strong magnetic field strengths $B\geq10^{15}$ G \citep{1992ApJ...392L...9D, 1993ApJ...408..194T, 1998PhRvL..81.4301D}. What leads to the formation of such magnetars is an open astrophysical question, but several channels have been proposed such as the mergers of binary neutron stars (BNSs) \citep{2013ApJ...771L..26G} or binary white dwarfs \citep{2013A&A...558A..39T,2015MNRAS.453.1910S}, core-collapse supernova associated with long gamma-ray bursts (LGRBs) \citep{Usov:1992zd,2000ApJ...537..810W, 2008MNRAS.383L..25B, 2009MNRAS.396.2038B} and accretion-induced collapse of white dwarfs (WDs) \citep{2013A&A...558A..39T}.  Although not yet ``seen'' directly, there are hints from the X-ray plateau of gamma-ray bursts (GRBs) that the central engine could likely be a rapidly rotating magnetar \citep{Rowlinson2010, Rowlinson2013, HaskellRavi, Lasky2017, Sarin2019}. Moreover, millisecond magnetars could explain the possible physics behind the observed plateaux in X-ray light curves of short gamma-ray bursts (SGRBs) \citep{Strang2019}. In all cases, the magnetar is born in an environment rich in matter, which facilitates accretion onto the star influencing its overall evolution. Such an object has a high rotational energy, which allows for the magnetar to be `supramassive', i.e. to support a higher maximum mass than a non-rotating star. As the star spins down due to gravitational wave (GW) and electromagnetic torques, this reduces centrifugal support and, unless the total mass of the system is low enough for the NS to be stable, eventually leads to collapse to a BH, as is generally expected for most binary NS merger remnants \citep{Ravi14}. However, in a matter rich environment, fallback accretion can play a leading role in the evolution of the system, as it provides a spinup torque which increases centrifugal support, but, ultimately, pushes the star closer to collapse to a BH by increasing its mass. This is particularly true if the magnetar is formed by the collapse of a massive star, in which case a massive disc of up to $\approx 1 M_\odot$ may be formed, leading to a hyperaccretion disc and powering part the emission observed in LGRBs \citep{MezRev}. A similar situation may occur after a binary merger, but the expelled mass that is subsequently re-accreted is expected to be lower, and not exceed $\approx 0.2 M_\odot$ \citep{Bernuzzi2020}.
In all cases, however, a transient source of GWs is expected, and it is essential to understand the early evolution of the system, and the impact of accretion, to determine the astrophysical relevance of these scenarios as targets for current and future GW detectors \citep{MuraseRev}. There have also been strong evidence based on the distribution of collapse times of millisecond magnetars that they spin-down through GWs among other things \citep{Sarin2020}.

After the first detection of GWs \citep{Abbott:2016blz}, the increasing rate of observing events -- compact binary coalescence including binary black holes, black hole-neutron star or BNSs -- by Advanced LIGO (aLIGO) and Advanced Virgo have opened new ways to look into the universe \citep{LIGOScientific:2018mvr, Abbott:2020khf, Abbott:2020uma}. GWs from isolated systems such as pulsars, newly-born magnetars or core-collapsed supernova still remain unobserved \citep{Collaboration:2009rfa,Pisarski:2019vxw,Abbott:2019heg, Abbott:2019prv,Authors:2019fue, LIGOScientific:2019hgc} demanding better theoretical models and improved sensitivity of the detectors. The distinctive signature of the GW strain and the rate of such events carry invaluable information about the properties of NSs such as its mass and radius \citep{Abbott:2018exr}. 

A number of mechanisms have been proposed by which an isolated NS can emit GWs \citep{Lasky:2015uia,2015ASSP...40...85H}. Firstly, all modes of oscillation can couple to the gravitational field, leading to GW emission. Following the birth of the star, the f-mode is the prime candidate to be excited and emit observable GWs  \citep{Ciolfi:2011xa}, however the r-mode oscillations in the NS core may be unstable if the star is born rapidly rotating, and contribute to its GW emission \citep{Owen1998,2001IJMPD..10..381A, 2007PhRvD..76f4019B,2009MNRAS.397.1464H,2014ApJ...781...26A,2015IJMPE..2441007H}. 

Secondly, a strong toroidal magnetic field could deform the shape of the star to a prolate-spheroid \citep{Cutler:2002nw} leading to unstable free precession and becoming an orthogonal rotator. \cite{DallOsso:2018dos} have shown that such an object with a spin period $\simeq 2 \, ms$ and an optimal ellipticity $\varepsilon \sim (1-5) \times 10^{-3}$ are potential candidates for aLIGO and future GW detectors. Additionally, in accreting NSs, the flow of matter onto the surface could lead to crustal asymmetries \citep{Ushomirsky2000} and create so called ``mountains'' when matter gets submerged deep within the crust \citep{2015MNRAS.450.2393H, Singh20, Gittins20}. The flow of matter also compresses the magnetic field both globally and locally which gives rise to a sizable mountain \citep{Payne2004,2005ApJ...623.1044M}. 

\begin{figure}
	\centering
	\includegraphics[scale=0.33]{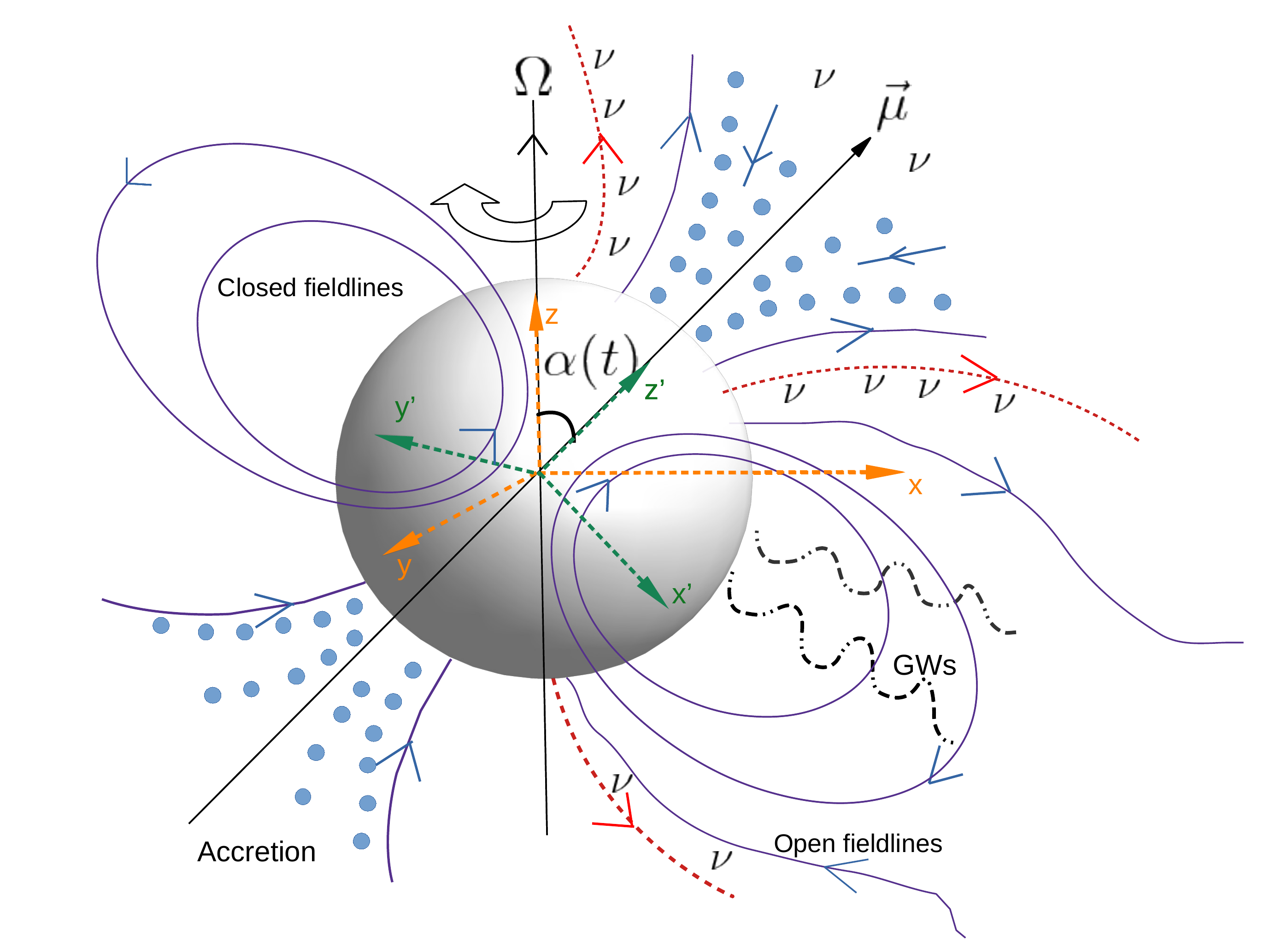}
	\caption{Pictorial description of the millisecond magnetar. There are two coordinate systems, one having the rotation axis ($\Omega$) and another having the magnetic moment ($\vec{\mu}$) axis. These axes are inclined at an angle $\alpha(t)$. The blue dots show matter falling on the two polar caps and forming two accreting columns. As the star rotates, it radiates energy in dipolar radiation and gravitational waves. The red dotted lines represent the escaping neutrinos which carry away heat and angular momentum in the form of a wind.}
	\label{NS}
\end{figure}

In this paper, we consider a newly born magnetar (figure \ref{NS}) rotating with millisecond period, formed in an environment where the matter around it could not reach its escape velocity and thus falls back. When the corotation radius exceeds the magnetospheric $\alfven$ radius, matter flows along the magnetic field lines and gives up angular momentum to the star. As the flow continues, two accreting columns are formed at the poles. In these regions, the freely falling material and the outflow reaches hydrostatic equilibrium \citep{1976MNRAS.175..395B} and neutrinos carry away heat and part of the gravitational binding energy. In these conditions the accretion rate ($\dot{M}$) in high, leading to super-Eddington accretion and significant accumulation of matter at the base of the columns, allowing the star to possess a time-varying quadrupole and emit GWs \cite{Piro12, 2019PhRvD.100l3014Z}. In this paper we study the spin evolution of the star, due to the different torques acting upon it, and calculate the characteristic GW strain. 

We improve upon the static model proposed in \cite{2019PhRvD.100l3014Z} by considering time-varying quantities such as the accretion rate, spin period, magnetic field and mass and radius, obtained from relativistic rotating stellar models, thus making the model truly \textit{dynamical}.  
We also consider an additional torque due to the neutrino driven wind of charged particles and show that it doesn't significantly spin-down the star. 
To further simplify our model we do not consider additional GW torques beyond those due to our `mountain' (e.g. due to unstable modes or hydromagnetic instabilities as in \cite{Melatos14}) nor do we consider viscosity in the stellar interior, as we shall see that this would impact the evolution of the system on timescales much longer than those of interest for our model.
With all these considerations, we show that accretion causes the star to spin faster radiating energy in GWs detectable by future observatories. 

This article is arranged as follows: we discuss the process of accretion leading to the formation of the massive NS and the mechanism by which it emits GW in sections 2 and 3 respectively. In section 4, we show our results for the spin evolution and GW strain for different initial conditions, while summary and conclusions are presented in section 5.

\begin{figure}
	\centering
	\includegraphics[scale=0.47]{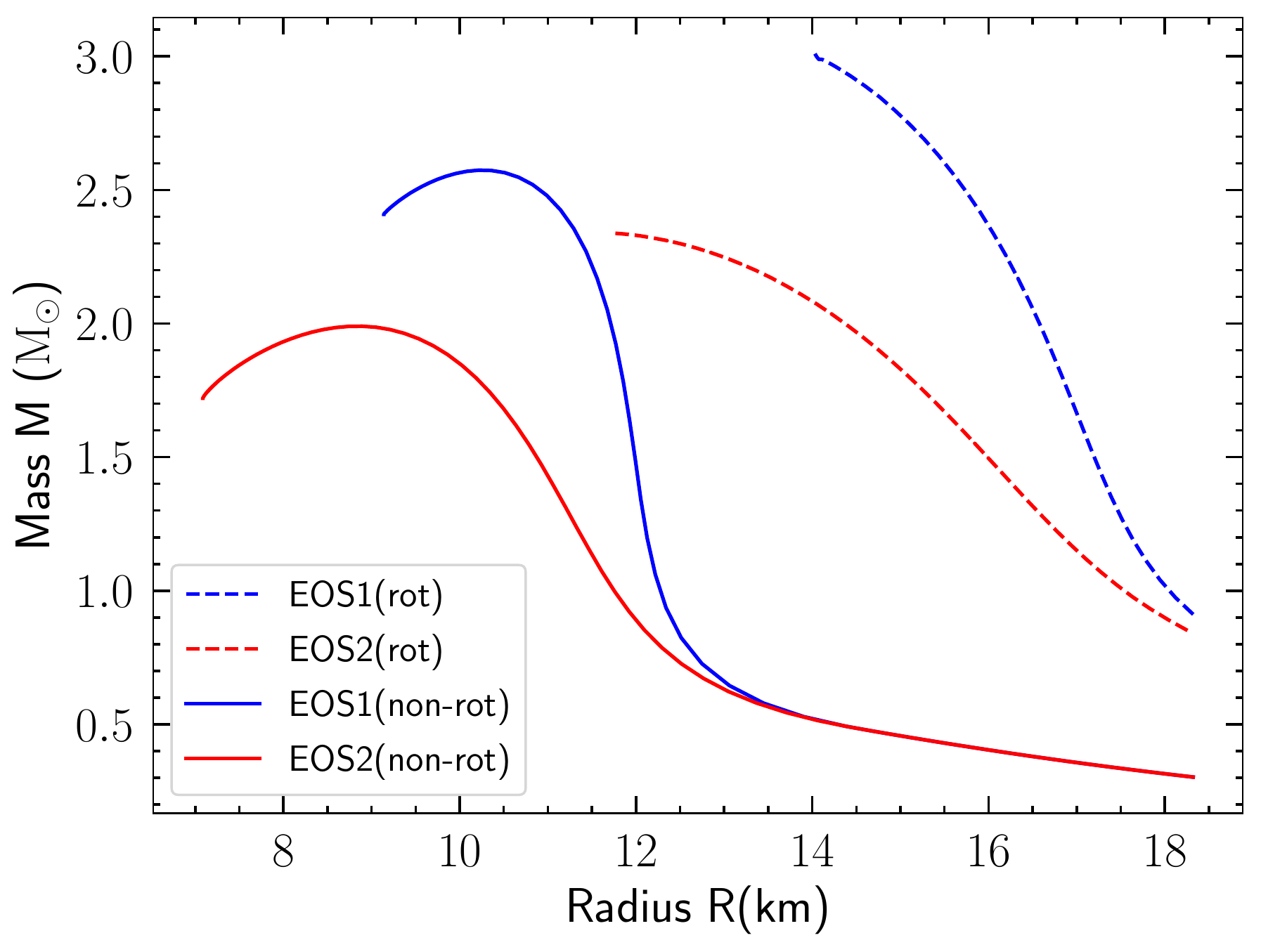}
	\caption{Mass-radius relation for our magnetar model using the two different 
          EOS given in table \ref{table:1}, i.e. blue solid line gives $M_{max}=2.57 \, M_{\odot}$ while the red solid line gives $M_{max}=2.0 \, M_{\odot}$. These refer to the
          non-rotating models and are obtained by solving the TOV
          equation. For a star rotating at Keplerian frequency, the
          gravitational mass and the equatorial radius are obtained
          with the RNS code, and plotted for the two different EOSs: higher maximum mass (blue dotted line) and the lower maximum mass (red dotted line).}
	\label{EOS}
\end{figure} 

\section{Accretion}
Two important radii that govern the flow of matter around a rotating NS are the magnetospheric $\alfven$ radius ($r_m$) and the corotation radius ($r_c$) defined as 
\begin{linenomath}
	\begin{align}
	&r_m = \bigg(\frac{B^4R^{12}}{GM\dot{M}^2}\bigg)^{1/7}, \,\,
	r_c = \bigg(\frac{GM}{\Omega^2}\bigg)^{1/3}
	\end{align}
\end{linenomath}
where $R$, $M$ and $\Omega$ are the radius, total mass and angular frequency of the magnetar. 
When $r_m > r_c$, matter spins at super-Keplerian speed and co-rotates
with the star. When $r_m < r_c$, the flow of matter gets channeled by
the magnetic field lines and accreted onto the two polar caps of the
NS, before spreading on the surface. We consider an
  early-type mass accretion rate $\dot{M} = 10^{-3} \eta t^{1/2}
  \,\,M_{\odot}s^{-1} $ which is obtained from fits following
  numerical simulations of supernova fallback accretion
  \citep{MacFayden2001, Zhang2008}. Here $\eta$ is a constant that
  depends on the supernova explosion process. Although the
  uncertainties remain large, and the exact scaling may be different
  than $t^{1/2}$, given the fact that the star accretes at a rate of
  $0.001-0.01 \, M_{\odot}s^{-1}$, this will not change significantly the survival time of the NS before collapsing to a BH. The total baryonic mass of the star as function of time (measured in seconds) is
\begin{linenomath}
	\begin{align}
	M_b(t) &= M_0 + \int_{0}^{t}\dot{M}dt \\
	&= M_0 + \int_{0}^{t}\eta 10^{-3} t^{1/2} dt \\
	&= M_0 + \frac{2}{3}\eta 10^{-3}t^{3/2}
	\end{align}
\end{linenomath}
where $M_0$ is its initial mass. In practice, to study the accretion torques, we will need the gravitational mass, which we denote with $M$. To obtain this we calculate, for each
time-step, a relativistic rotating model for the star, using the code \textsc{RNS} \citep{RNS}, and extract the gravitational mass, and moment of inertia of the star, as we shall describe in the following.  As the magnetar gains mass (when $r_m < r_c$), its radius changes depending on the equation of state (EOS) of the NS. This requires detailed modeling of the accreted crust \citep{HaenselBook, Haensel2008, Gusakov:2020}. Since we are interested in the approximate behaviour of how the radius changes with
mass, we adopt a simple EOS of the form
\begin{linenomath}
\begin{align}
	P(\rho) =
	\begin{cases}
	 k\rho^{\gamma_1}     & \quad \text{if } \rho<\rho_c \\
	k^{\prime}\rho^{\gamma_2} & \quad \text{if }  \rho \geq \rho_c
	\end{cases}
	\label{EOSeq}
\end{align}
\end{linenomath}
where $\gamma_1=1.663$ considering non-relativistic neutrons in the outer layers (leaving the inner layers unaffected), $\gamma_2=3.4$, and $\rho_c = 4.5\times10^{14} \gcc$. We fixed $k = 3^{2/3}\pi^{4/3}h^2/5m_n^{8/3}$, where $m_n$ is the mass of a neutron while the constant $k^{\prime}$ was calculated imposing continuity of pressure at the critical density $\rho_c$. This particular EOS was adopted since it models the physics of the outer layers of the
star, which are affected by the accretion and determine the change in radius, but allows us to obtain a $1.4 M_{\odot}$ star with radius $R=12$ km for the non-rotating model, is consistent with more realistic estimates obtained from micro-physically motivated equations of state
\citep{HaenselBook}. The mass radius relation for this EoS, both in the case of a non-rotating star, obtained by solving the Tolman-Oppenheimer-Volkoff (TOV) equation, and for a model rotating at the Keplarian breakup frequency (obtained using RNS) are shown in figure \ref{EOS}.
The maximum mass obtained for the non-rotating case is $2.57 M_{\odot}$, interestingly close to the recent observation by LIGO of a possibly heavy NS \citep{Abbott:2020khf}. We assume that the star collapses to a BH when this mass limit is exceeded. Additionally, as can be seen in figure \ref{EOS} we explored an EOS by using the same $
\gamma_1$ and $\rho_c$ but changing $\gamma_2 = 2.65$, yielding a
maximum mass $M_{max} = 2 M_{\odot}$, which is consistent with the maximum observed mass of a NS to date
\citep{2010Natur.467.1081D, Cromartie2020}. Table \ref{table:1} shows the summary of the parameters and the maximum non-rotating mass for both the EOS considered in our analysis. The accretion column forms when $r_m < r_c$ implying a critical accretion rate 
\begin{linenomath}
	\begin{align}
	\dot{M} > \dot{M}_{cric} = 1.8 \times 10^{-2} M_{1.4}^{-5/3} B_{15}^{2}R_{12}^{6}P_1^{-7/3} M_{\odot} s^{-1}
	\end{align}
\end{linenomath}
showing a strong scaling with the radius and spin period of the star. 

\begin{center}
\begin{table}
	\centering
	\begin{tabular}{||c c c c c||} 
		\hline
		Equation of state & $\gamma_1$ & $\gamma_2$ & $\rho_c$(\gcc)& M$_{max}^{non-rot} (M_{\odot})$ \\ [0.5ex] 
		\hline\hline
		EOS1 & 1.663 & 3.4 &  $4.5\times10^{14}$ &2.57\\ 
		\hline
		EOS2 & 1.663 & 2.65 & $4.5\times10^{14}$ &2.0 \\
		\hline
	\end{tabular}
	\caption{Summary of the parameters considered for the two different equation of states for a non-rotating NS.}
	\label{table:1}
\end{table}
\end{center}

\section{Gravitational Waves}
Let us consider the magnetar to be a rigid-body rotating with an angular velocity $\Omega$ about the $z$-axis (see figure
\ref{NS}). The magnetic axis points along $z'$ axis, inclined at an angle $\alpha (t)$ with respect to the rotation axis. 

We can express the  moment of inertia of a spherical star, as $I = \tilde{I}MR^2$, where the behaviour of $\tilde{I}$ is depends on the equation of state, but at least in slow rotation is generally a function of compactness \citep{LattimerSchutz, Breu16}.

As our star is not spherical, but deformed by the presence of the
accretion mound, let us denote the moment of inertia along the
coordinate axes ($x^{\prime},y^{\prime},z^{\prime}$) as $I_1$, $I_2$
and $I_3$. The system won't be exactly biaxial as the
  rotational bulge is associated with the x, y, z coordinate system,
  whereas the accretion mountain is associated with the x', y', z'
  coordinate system. Given the many other approximations of our
  analysis, we will however ignore this small effect and take $I_1 \sim I_2$. The GWs emitted by such an object have an amplitude
\begin{linenomath}
	\begin{equation}
	h_0 = \frac{4G}{c^2}\frac{(I_1-I_3)\Omega^2\sin^2{\alpha}}{d}
	\end{equation} 
\end{linenomath}
where $d$ is the distance to the source. If we imagine the accreted matter to be two cylindrical bodies with radius $r$ at the poles, the moment of inertia along the $x^{\prime},z^{\prime}$ axes follows as
\begin{linenomath}
	\begin{align}
	&I_1 = \tilde{I}MR^2 + 2M_{acc}R^2\\
	&I_3 = \tilde{I}MR^2 + 2\times \frac{1}{2}M_{acc}r^2
	\end{align}
\end{linenomath}
where $M_{acc}$ is the mass accreted at each pole. The difference in moment of inertia along the two directions can be approximated as
\begin{linenomath}
	\begin{align}
	I_1-I_3 &= 2M_{acc}R^2 - 2 \times \frac{1}{2}M_{acc}r^2\\
	&\approx 2M_{acc}R^2, \, \,  \, R \gg r
	\end{align}
\end{linenomath}
Thus the GW amplitude becomes 
\begin{equation}
h_0 = \frac{8G}{c^2}\frac{M_{acc}R^2\Omega^2\sin^2{\alpha}}{d}
\end{equation} 
and the characteristics GW strain is approximately given by \citep{Corsi2009}
\begin{linenomath}
	\begin{align}
	h_c = fh_0\sqrt{\frac{dt_{sur}}{df}} \approx h_0\sqrt{ft_{sur}}
	\end{align}
\end{linenomath}
where $t_{sur}$ is the survival timescale for the magnetar before its collapse, $f=\Omega/\pi$ is the dominant frequency at which GWs are emitted for
$\alpha = 90^{\circ}$. The body does not emit any GWs when $\alpha$
either becomes $0$ or $\pi$ radians. The gravitational wave luminosity in such a process goes as
\begin{linenomath}
	\begin{align}
	L_{GW} &= - \frac{2}{5}\frac{G}{c^5}(I_1-I_3)^2\Omega^6\sin^2\alpha  (16\sin^2\alpha+\cos^2\alpha)\\
	&= -\frac{8}{5}\frac{G}{c^5}M^2_{acc}R^4\Omega^6\sin^2\alpha  (16\sin^2\alpha+\cos^2\alpha)
	\end{align}	
\end{linenomath}

\section{Spin and Inclination}
The spin evolution is affected by the torques acting due to accretion, the emission of gravitational waves, the escaping neutrinos and the dipolar magnetic field radiation. The rate at which angular momentum is lost due to the GWs is 
\begin{linenomath}
	\begin{equation}
	N_{GW} = \frac{L_{GW}}{\Omega}
	\end{equation}
\end{linenomath}
For the torque due to the external dipolar magnetic field we use the
expression deduced by \cite{Spitkovsky:2006np} from numerical
simulations of the magnetosphere in plasma:
\begin{linenomath}
	\begin{align}
	N_{dip} = \frac{B^2R^6\Omega^3}{c^3}(1+\sin^2\alpha)
	\end{align}
\end{linenomath}
The accretion torque acting on the magnetar is given by
\begin{linenomath}
\begin{equation}
N_{acc} = n(\omega)(GMr_m)^{1/2}\dot{M}
\end{equation}
\end{linenomath}
where we adopt $n(\omega) = (1-\omega)$ as considered by \cite{Piro,2019PhRvD.100l3014Z}. This torque  can either be positive or negative depending on the fastness parameter $\omega=(r_m/ r_c)^{3/2}$. The star enters the so-called ``propeller phase'' for $n(\omega) <0$ where it experiences a negative torque and spins down by expelling matter from the super-Keplerian magnetosphere. In fact, \cite{Eksi2005} showed that in most cases of fallback accretion, the disk will pass through this propeller phase and such systems could appear as Ultra-Luminous X-ray Sources when the disk is fed by super-critical mass accretion rates \citep{Erkut2019}. For a detailed discussion, see \cite{Zhang2008, Piro, Dai2012}. \\

The neutrino driven wind is expected to interact with the strong magnetic field of the star, leading to co-rotation of charged particles in the magnetosphere and a loss of angular momentum \citep{Thompson:2004wi}. This emission, in fact, follows the open magnetic field lines and is thus not isotropic. The luminosity and the energy of neutrinos in this process are given by \citep{2020MNRAS.494.4838L}:

\begin{linenomath}
	\begin{equation}
	\frac{L_{\nu}(t)}{10^{52} \,\rm erg \, s^{-1}} = 0.7\exp\bigg(-\frac{t}{1.5 s}\bigg) + 0.3\bigg(1-\frac{t}{50 s}\bigg)^{4} 
	\end{equation}
	\begin{equation}
	\frac{E_{\nu}(t)}{10 \,\rm MeV} = 0.3\exp\bigg(-\frac{t}{4 s}\bigg) + \bigg(1-\frac{t}{60 s}\bigg) 
	\end{equation}
\end{linenomath}
These fits have been obtained from the simulations by \cite{Pons1999,
  Mertzger2011} and are valid upto $40 \,s$. Although these
expressions are valid for slow-rotation, \cite{2020MNRAS.494.4838L}
makes improvements by considering centrifugal enhancement for which we
use the limiting values. Based on our reference model,
  with the uncertainties, we may expect different behaviour due to
  also different temperatures in the cases of core-collapsed
  supernovae and BNS post-mergers. But given that we confirm this
  torque is generally negligible for our problem, we do not investigate the effect of temperature further. The mass loss rate due to the neutrinos is given by:

\begin{equation}
\dot{M}_{\nu} = 6.8\times 10^{-5} \, \rm M_{\odot}s^{-1}\bigg(\frac{L_{\nu}}{10^{52} \,\rm \ergs}\bigg)^{5/3} \bigg(\frac{E_{\nu}}{10 \,MeV}\bigg)^{10/3}
\end{equation}
The  rate of change of electromagnetic energy carried away by the neutrinos is given by \citep{2020MNRAS.494.4838L}:
\begin{align}
&\dot{E}_{\rm EM} =
\begin{cases}
c^2\dot{M}_{\nu}\sigma_0^{2/3}      & \quad \text{if } \sigma_0 <1 \\
\frac{2}{3}c^2\dot{M}_{\nu}\sigma_0 & \quad \text{if } \sigma_0 \geq 1
\end{cases}
\label{alphaeq}
\end{align}
where $\sigma_0$ is called the wind magnetization parameter \citep{Mertzger2011} given by
\begin{linenomath}
	\begin{equation}
	\sigma_0 = \frac{B^2R^4\Omega^2}{\dot{M_{\nu}}c^3}
	\end{equation}
\end{linenomath}
that accounts for asymptotic partition between the kinetic and magnetic energy in the wind. Further, $\sigma_0 \leq 1$ implies a non-relativistic wind as compared to $\sigma_0 > 1$ for a relativistic wind \citep{Mertzger2011}. Neglecting the small change in moment of inertia of the star due to neutrino mass loss, the spin evolution purely due to the loss of neutrinos is given by

\begin{equation}
\frac{d {\Omega}}{d t} = -\frac{\dot{E}_{\rm EM}}{I\Omega} = -\frac{L_{\nu}}{I\Omega}
\end{equation}
Thus, the exact scaling with which $\Omega$ varies with time is
proportional to $\Omega^{1/3}$ for $\sigma_0<1$ while $\Omega$ for
$\sigma_0>1$. As in past studies, which have suggested that neutrinos
are inefficient in spinning down a protoneutron star
\citep{Baumgarte_1998}, we find that magnetic dipole radiation is a dominant effect as found by \cite{Lasky2017} leading to a braking index of $n=3$ obtained from fits of X-ray plateaux in SGRBs, although it's precise observational determination remains challenging \citep{Melatos97, Archibald2016}. \\

The net rate of change of angular momentum due to the different torques acting on our star is
\begin{equation}
\frac{d}{dt}(I_3\Omega) = -N_{GW} - N_{dip} - N_{\nu} + N_{acc} = N^{net}
\end{equation}
$I_3 \approx \tilde{I}MR^2$ (neglecting the contribution from the cylindrical column of matter). We drop the notation $I_3$ and simply represent it as $I$. This gives us an evolution equation for the spin,
\begin{align}
&\frac{d \Omega}{dt} = (-N_{GW} - N_{dip} - N_{\nu} + N_{acc})/I - \frac{\Omega}{I}\frac{dI}{dt}
\end{align}
and
\begin{equation}
\frac{1}{I}\frac{dI}{dt} = \frac{\dot{M}}{M} + 2\frac{\dot{R}}{R} + \frac{\dot{\tilde{I}}}{\tilde{I}}
\end{equation}
where $\dot{M} = \dot{M}_{acc} - \dot{M}_{\nu}$ and the dot represents derivative with respect to time. The mass loss due to neutrinos allows us to calculate the evolution of inclination angle (for a more sophisticated treatment including the effect of internal dissipation due to viscosity, see \cite{2020MNRAS.494.4838L}) as follows
\begin{linenomath}
	\begin{align}
	&\frac{d\alpha}{dt} = \frac{\dot{E}_{EM}\sin\alpha\cos\alpha}{I\Omega^2} 
	\end{align}
\end{linenomath}
Viscosity plays an important role in affecting the inclination angle of NSs \citep{1976Ap&SS..45..369J,2017MNRAS.472.2142D}. The compelling effects due to viscous dissipation and external torques tend to make the magnetic axis orthogonal to the rotational axis soon after birth which gradually starts aligning over hundreds of years \citep{2020MNRAS.494.4838L}. Since we are interested in shorter timescales during which the star survives, we ignore the evolution of $\alpha(t)$ due to bulk viscosity and dissipation from the internal fluid motions.

\vspace{0.4 cm}
\section{Results\label{results}}
\label{results}
\subsection{Core-collapsed supernovae (CCSNe)}
\begin{figure}
	\centering
	\includegraphics[scale=0.47]{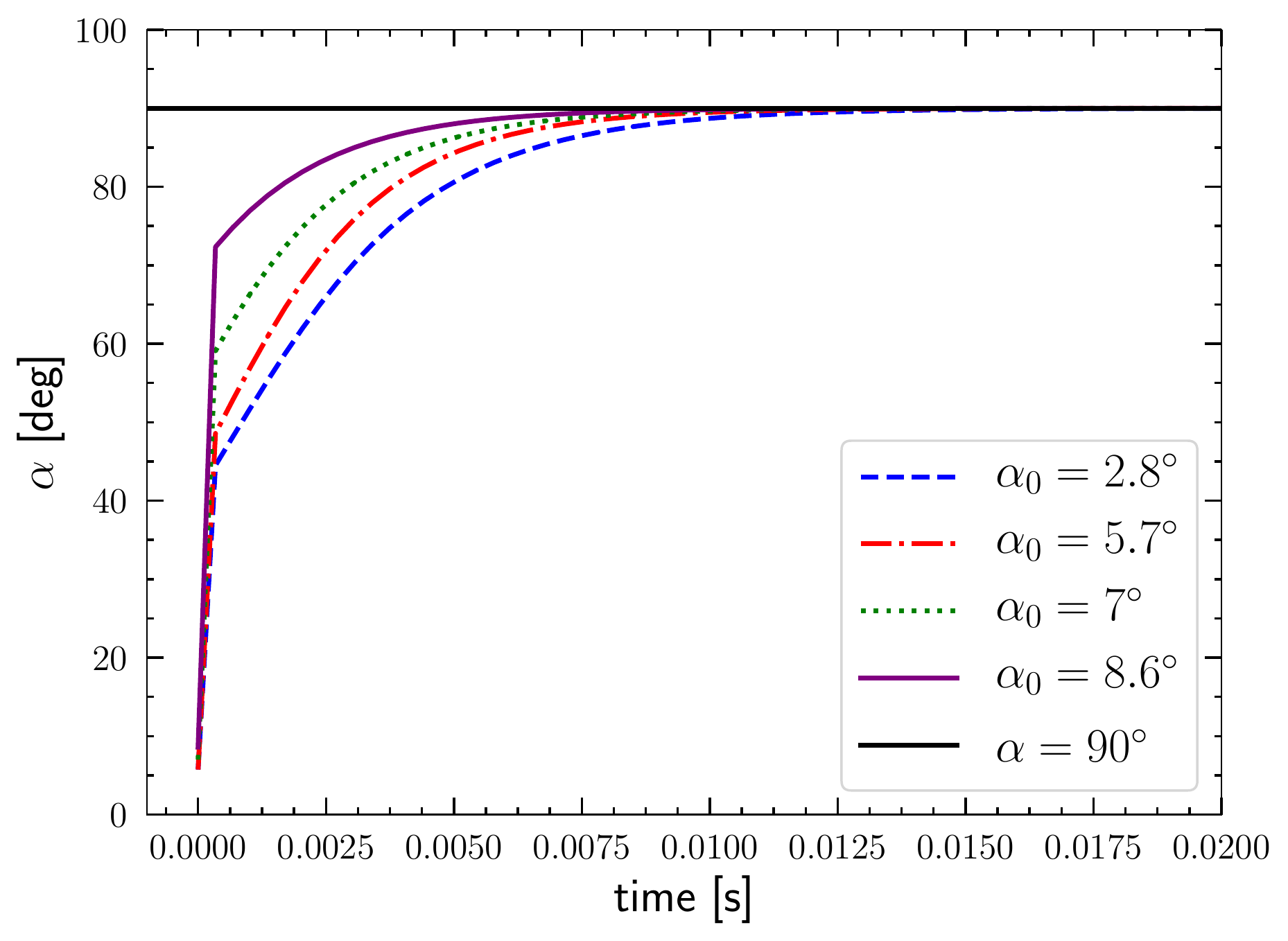}
	\caption[short]{Evolution of the inclination angle $\alpha(t)$ for the magnetar formed after CCSNe for initial values $\alpha_0 \in ( 2.8^{\circ} , 5.7^{\circ}, 7.0^{\circ}, 8.6^{\circ}$). The black solid line denotes $\alpha=90^{\circ}$. Given any initial choice, $\alpha \to 90^{\circ}$ in $t \leq 10$ ms.}
	\label{alpha}
\end{figure}

\begin{figure}
	\centering
	\includegraphics[scale=0.53]{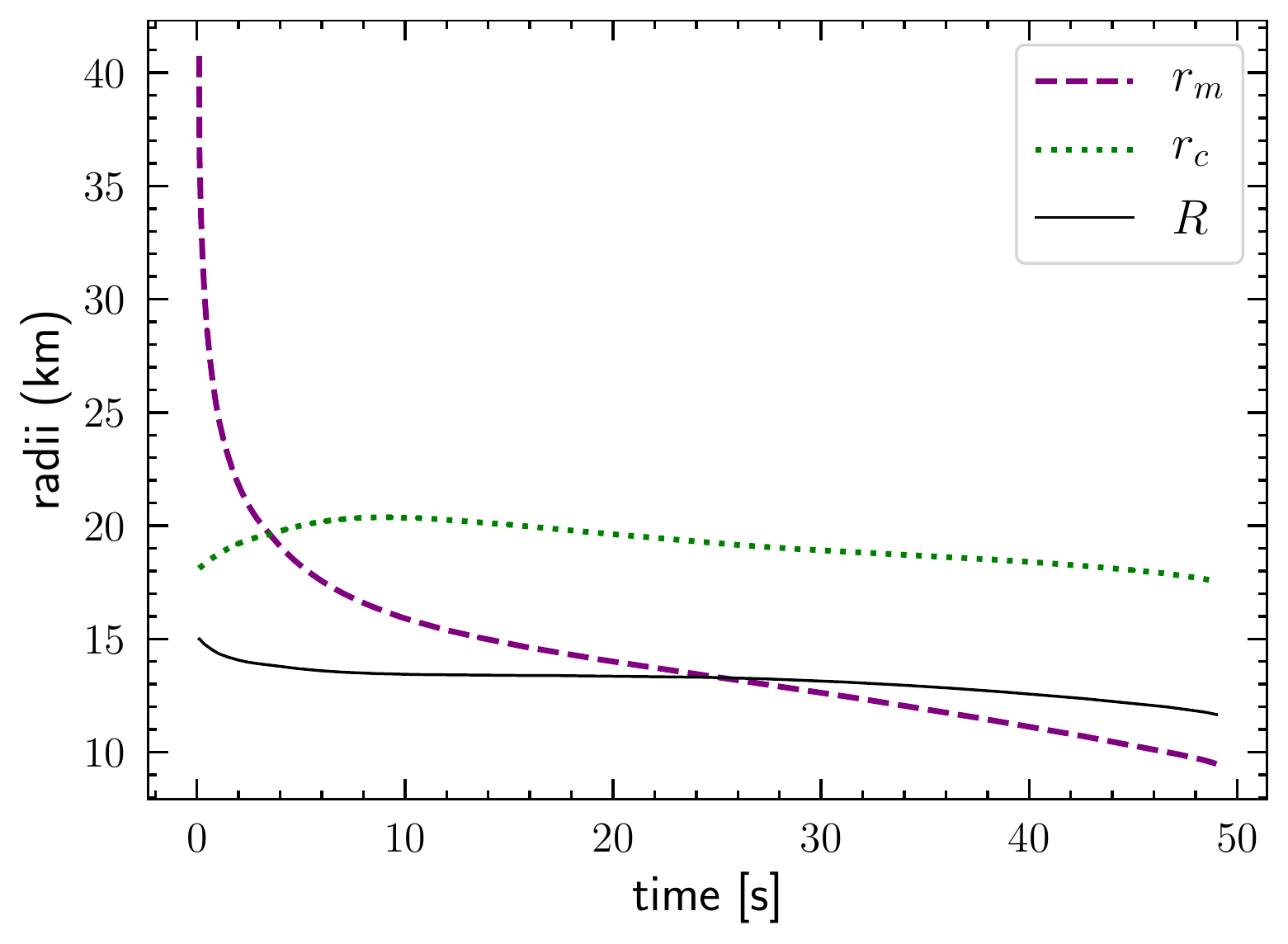}
	\caption[short]{The location of magnetospheric radius ($r_m$), corotation radius ($r_c$) and the radius of the star ($R$) as a function of time for the magnetar formed after CCSNe with initial mass $M_0=1.4 M_{\odot}$ and $P_0=1.1$ ms. When $r_m$ becomes less than $R$, we set $r_m=R$ in our simulation.}
	\label{rloc}
\end{figure}

\begin{figure}
	\centering
	\includegraphics[scale=0.35]{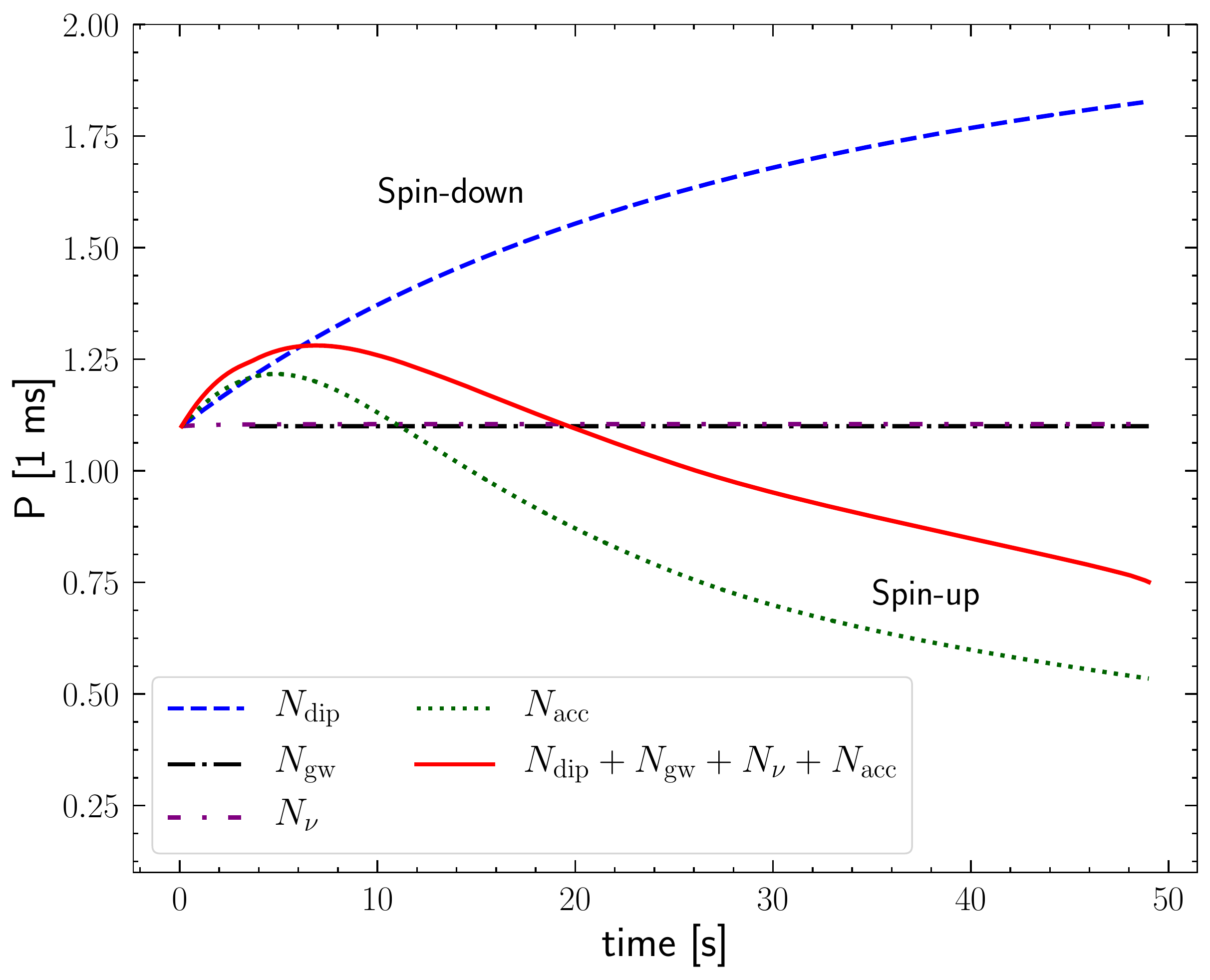}
	\caption{The spin evolution of the magnetar formed after CCSNe due to the various individual torques and the overall behaviour represented by solid red line. The period initially increases and then starts decreasing with time as accretion wins over the combined effects of neutrino luminosity, dipolar radiation and the GW luminosity. At the end, the star collapses to a BH. }
	\label{spin}
\end{figure}

We present our results after the birth of the magnetar with EOS1. The value of $\eta$ plays an important role in the lifetime before the magnetar collapses to a BH. Simulation shows that typical values of $\eta$ lie between $0.1-10$ \citep{Piro}. A lower $\eta$ implies a more powerful supernova explosion and varying this for the magnetar, we expect a change in $t_{sur}$, the rate at which it gains mass and the duration of the GW signal. However, this should not affect the overall shape of the waveform.

We chose $\eta=10$ making the magnetar (with $M_0 = 1.4 M_{\odot}$ and $P_0=1.1$ ms) to survive a total of $t_{sur} \sim 49 \,s$. The choice of these parameters has several implications, first being the time at which accretion columns are formed as one requires $\dot{M} > \dot{M}_{cric}$ \citep{2019PhRvD.100l3014Z}. This condition is achieved at a threshold time $t_{th} \sim 2.0 \,s$. Second, $n(\omega) < 0$ for $t < t_{th}$ and the star experiences a negative torque. Third, there are no GWs emitted during this phase. The spin evolution is only affected by the dipolar radiation, angular momentum loss due to neutrinos and the co-rotating matter. After $t>t_{th}$, two accretion columns are formed, the star experiences a positive torque and the GWs carry away angular momentum.
Since it is easier for us to work in normalized units, we express different quantities of the star such as mass, radius, magnetic field, and the mass accretion rate as
\begin{linenomath}
	\begin{equation}
	R_{15} = \frac{R}{15 \,\rm km} ,\, M_{1.4} = \frac{M}{1.4 M_{\odot}}, \, B_{15} = \frac{B}{10^{15} G}, \, \dot{M}_{-2} = \frac{\dot{M}}{10^{-2} M_{\odot}s^{-1}}
	\label{scaling}
	\end{equation}
\end{linenomath}
We further work with spin period ($P = 2\pi/\Omega$) which is normalized as $P_1 = P/1 \,ms$. To calculate the amount of column mass that is accreted on to the poles, we use the relation obtained by \cite{2019PhRvD.100l3014Z}
\begin{linenomath}
	\begin{equation}
	M_{acc} = 1.7\times 10^{-7} M_{\odot} M_{1.4}^{-25/56} \dot{M}_2^{3/28}B_{15}^{-5/7}R_{15}^{125/56}
	\end{equation}
\end{linenomath}
This estimate is model dependent, however different estimates, e.g. considering magnetically confined matter \citep{Brown98, Singh20}, produce similar estimates for the mass. The scalings with the different parameters (i.e $M_{1.4}$, $\dot{M}_{-2}$, $B_{15}$, etc.) will be different and may affect our model. However the strongest driver of the evolution is the scaling of the GW torque with spin period, which will remain unaffected, as will the electromagnetic torques. We thus do not expect these uncertainties to affect our qualitative conclusions.
We assume that the accretion column reduces the magnetic dipole moment as $\mid \mu \mid = \mid \mu_i \mid (1-M_{acc}/M_c)$, with $M_c = 1.2\times10^{-6} M_{\odot}s^{-1}$ \citep{Payne2004}. This makes the magnetic field strength to vary as 
\begin{equation}
B_{15} = B_{15,i}\frac{R^3_{15, i}}{R^3_{15}}\big(1-\frac{M_{acc}}{M_c}\big)
\end{equation}
where $i$ denote the initial value of each quantity. The accreted material drags the magnetic field lines by flux freezing as it moves from the polar caps towards the equator. A detailed calculation of the magnetic field structure and density of the mountain requires us to solve the Grad-Shafranov equation for magnetic equilibria, which is beyond the scope of this paper, but is outlined in the work by \cite{Melatos14}. The field burial further facilitates the formation of BHs by shutting the propeller effect and allowing fallback accretion. The evolution of inclination angle can be calculated using equation \ref{alphaeq}. Figure \ref{alpha} shows that for any random initial choice of $\alpha$, the magnetic axis always becomes perpendicular to the rotation axis in about $10 \,ms$. The kinks present at early times are an artifact of grid resoluton. We choose the initial value of $\alpha = 5.7^{\circ}$ and our evolution at later stages is independent of this choice. By the time GWs are emitted, $\alpha$ becomes $90^{\circ}$ and emission reaches its peak value However, it is expected that in the first day of the magnetar, the inclination angle decreases rapidly making the rotation and magnetic axis aligned to each other \citep{Samaz2019,Samaz2020}. Using the definitions in equation \ref{scaling}, we find expression for the GW luminosity
\begin{linenomath}
	\begin{align}
	&L_{gw} = 1.1\times 10^{42} \ergs M_{1.4}^{-25/28} \dot{M}_2^{3/14}B_{15}^{-10/7}R_{15}^{237/28}P_{1}^{-6}
	\end{align}
\end{linenomath}
which changes with time as the magnetar's spin evolves due to the various processes. We calculate the evolution of the system accretes, and calculate sequences of fixed baryon mass models with RNS, to obtain at each time-step the gravitational mass, radius and moment of inertia of the star. We investigate a range of models with initial spin period above the Keplerian breakup period corresponding to the initial mass, which we calculate with the RNS code, and take as our reference model a magnetar with initial spin period $P_0 = 1.1 \, ms$, as in  \citet{Ott2006}.

\begin{figure}
	\centering
	\includegraphics[scale=0.48]{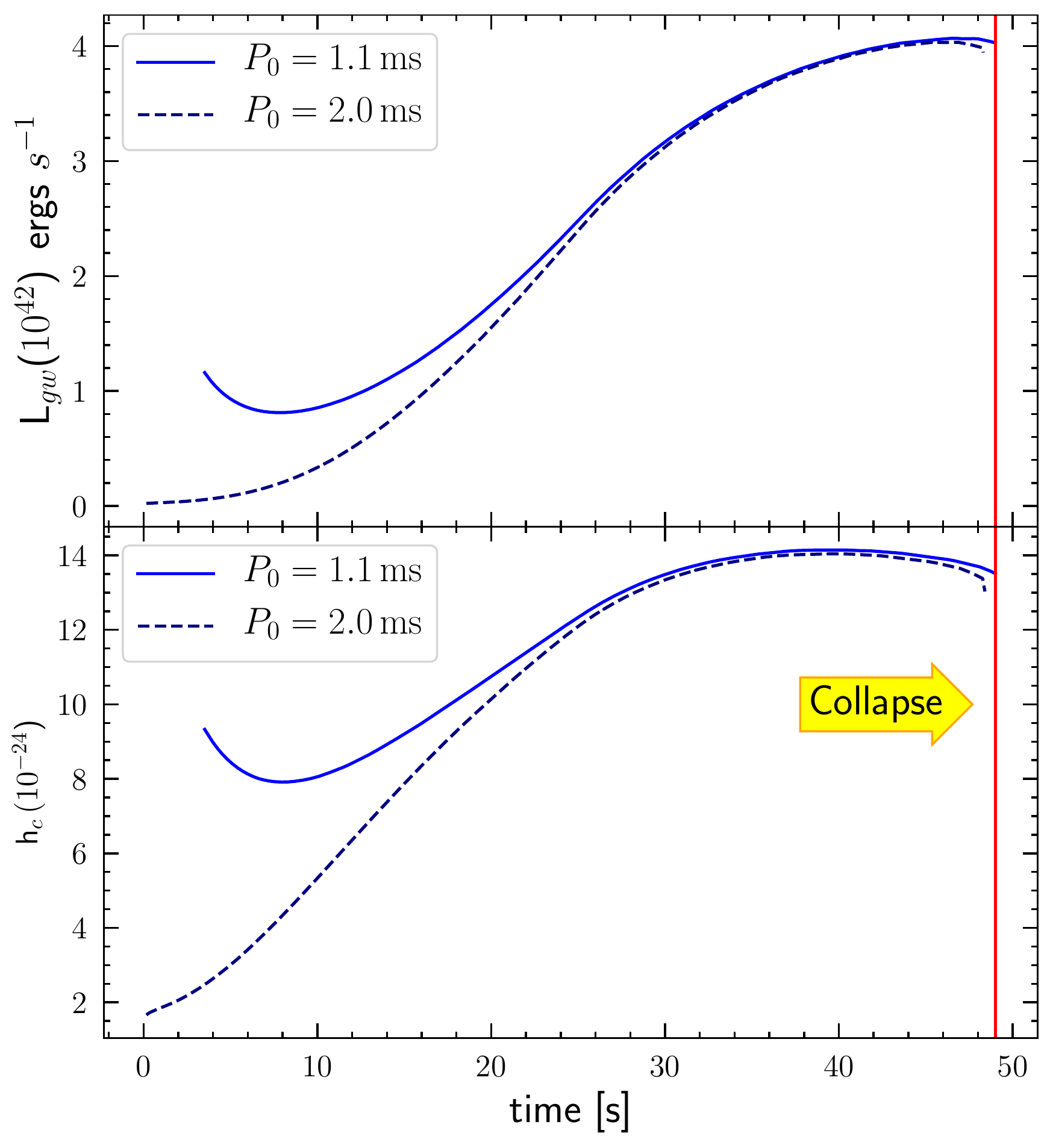}
	\caption{The GW luminsoity (top) and characteristic GW strain (bottom) as a function of time for two different initial spin periods of the magnetar formed from CCSNe. The red line shows the time at which the magnetar collapses to a BH and the GW signal truncates.}
	\label{gwlum}
\end{figure}

\begin{figure}
	\centering
	\includegraphics[scale=0.37]{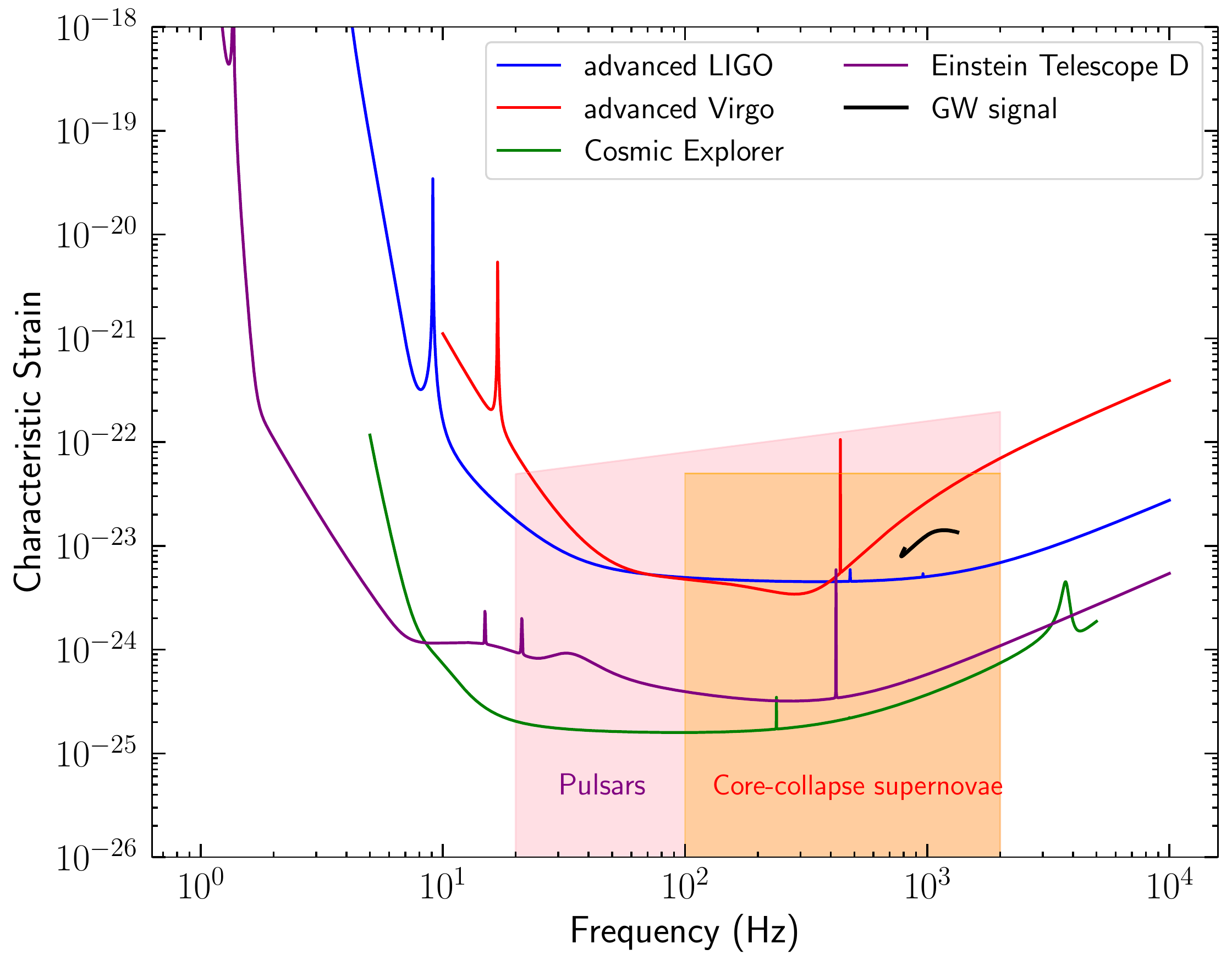}
	\caption{Design sensitivities for the second and third generation GW detectors to the characteristic strain. The purple, green, blue and the red lines represent the Einstein Telescope, Cosmic Explorer, advanced LIGO and advanced Virgo respectively. The black line represent our GW signal from the magnetar before its collapse to a BH. Also shown are populations of pulsars (pink) and core-collapse supernovae (orange) with their spin frequencies.}
	\label{SNR}
\end{figure}

\begin{figure}
	\centering
	\includegraphics[scale=0.55]{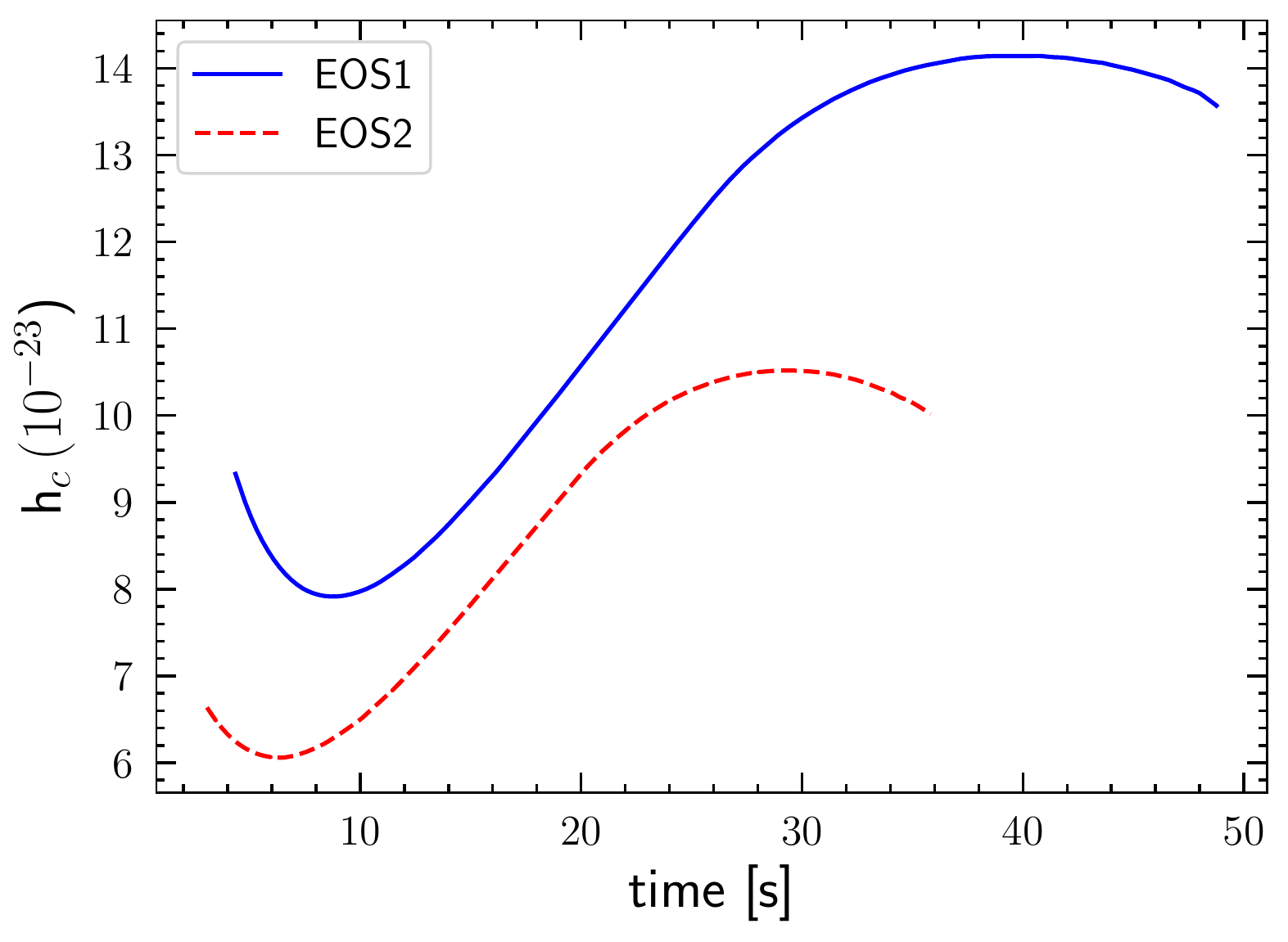}
	\caption{Characteristic GW strain for the two different equation of states (EOSs) in the case of CCSNe magnetar. Qualitatively the behaviour is same apart from the total duration of the signal and its peak value.}
	\label{hEOS}
\end{figure}

\begin{figure}
	\centering
	\includegraphics[scale=0.55]{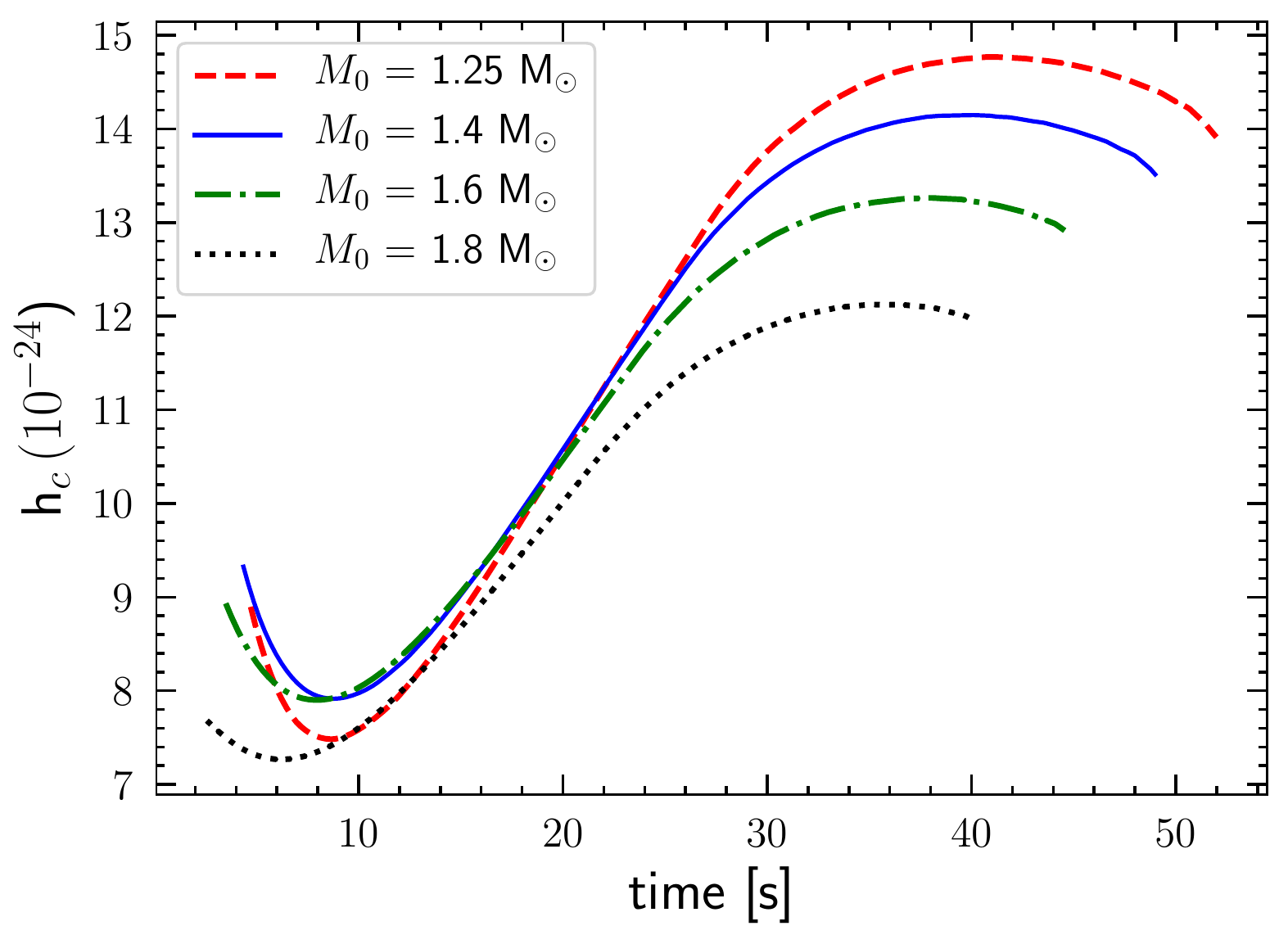}
	\caption{Characteristic strain for different remnant masses of the magnetar (formed from CCSNe), $M_0 = ({1.25,1,4,1.6,1.8}) \, M_{\odot}$. A lower mass progenitor emits GW strain at a higher strength before its collapse and has longer duration signal.}
	\label{h_varmass}
\end{figure}

Figure \ref{spin} shows the evolution of spin due to the various torques. We see that the magnetic dipole radiation carry away most of the angular momentum as compared to the neutrino wind and GWs, which results in an initial increase in the spin period. After $t>t_{th}$, the torque on the star due to accretion is positive and dominates over other processes. At this stage, the condition $r_c>r_m$ also remains true. However, when $r_m$ becomes less than the radius of the star, we set $r_m=R$. The evolution of these various radii is shown in figure \ref{rloc}, while in figure \ref{spin} we plot the contributions of the different torques to the spin evolution of the star. Similar results are also presented in \cite{Melatos14} where the evolution of angular frequency vs time is calculated for accretion mountain with and without GWs due to hydromagnetic instabilities. In figure \ref{gwlum} we show the GW luminosity and the characteristic strain for two different choices $P_0$. There is an initial decrease in both quantities, which then gradually start increasing with time. Moreover, a higher initial spin simply makes the star to emit GWs at a lower luminosity before the collapse. Note that this behaviour is obtained considering that the spin period and frequency are time-dependent unlike the results given in \cite{2019PhRvD.100l3014Z} which shows a rise in $h_c$ and $L_{gw}$ till $4.3 \,s$ and a fall afterwards with time. We stress that our model includes a fully relativistic rotating stellar model, and full time evolution of the system's parameters. Our model is thus more realistic and accurately represents the GW signal expected from this source, shown by the black line at around $\sim 10^3$ Hz in figure \ref{SNR}. On performing the same analysis for a NS with $M_{max} = 2 M_{\odot}$, we find a survival time of $35 \,s$ and no qualitative change in our results, see for example the behaviour of the GW strain in figure \ref{hEOS} . Furthermore, we have varied the initial mass of the NS from $1.25-1.8 \, M_{\odot}$. From figure \ref{h_varmass} we can see that lowering the mass produces a higher peak of $h_c$ and a longer signal.

\begin{figure}
	\centering
	\includegraphics[scale=0.36]{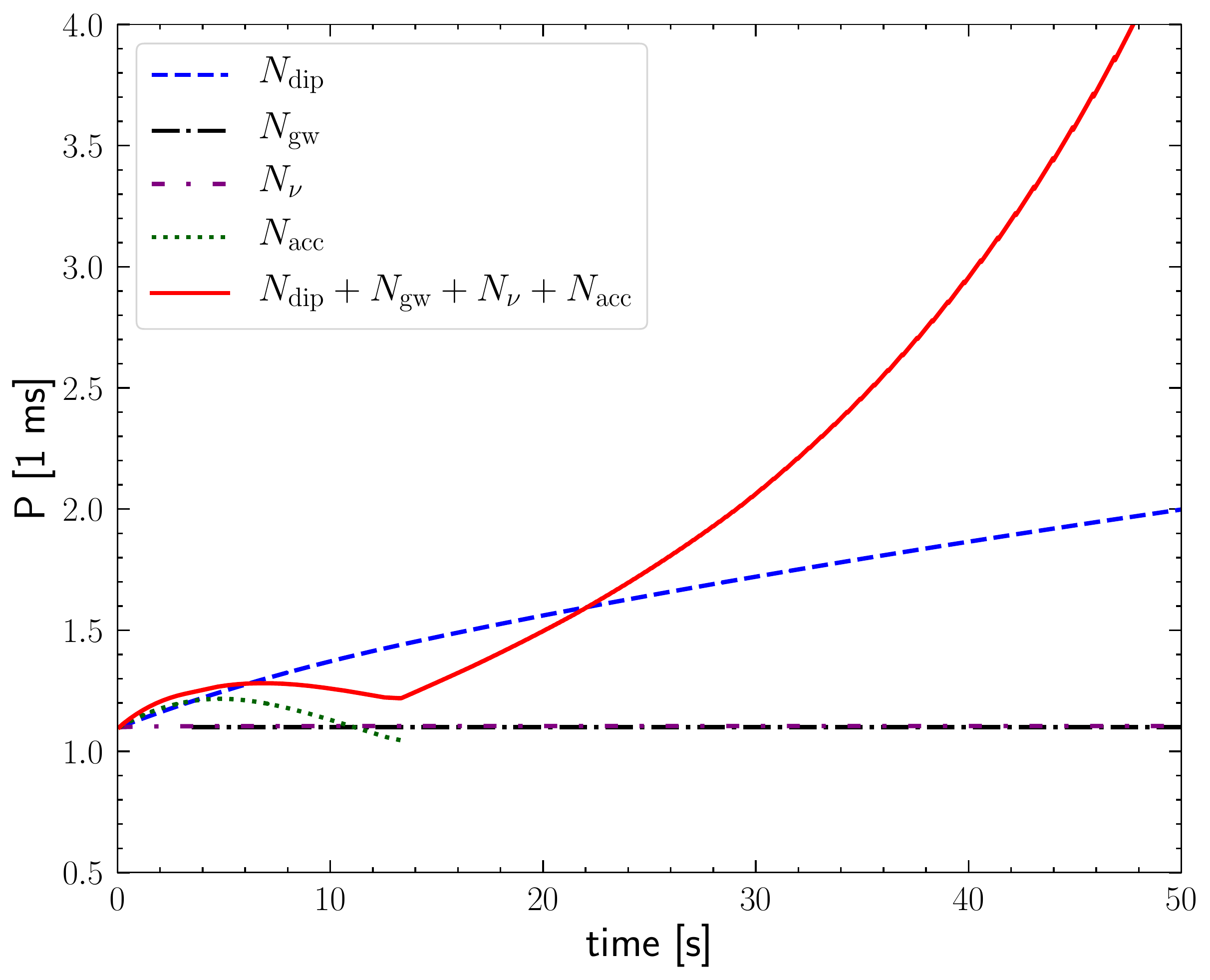}
	\caption{Spin evolution of the magnetar formed after the merger of a BNS (for a model with $\eta=10$) when the total mass available for accretion is $0.2 M_{\odot}$. The different lines represent the spin due to the various individual torques while the red solid line shows the overall effect on the spin. The accretion torque stops at 14 $s$ when the star starts to spin-down due to angular momentum carried away by the neutrino-wind, magnetic dipole radiation and emission of GWs.}
	\label{spin2}
\end{figure}

\begin{figure*}
	\centering
	\begin{subfigure}{.5\textwidth}
		\centering
		\includegraphics[scale=0.55]{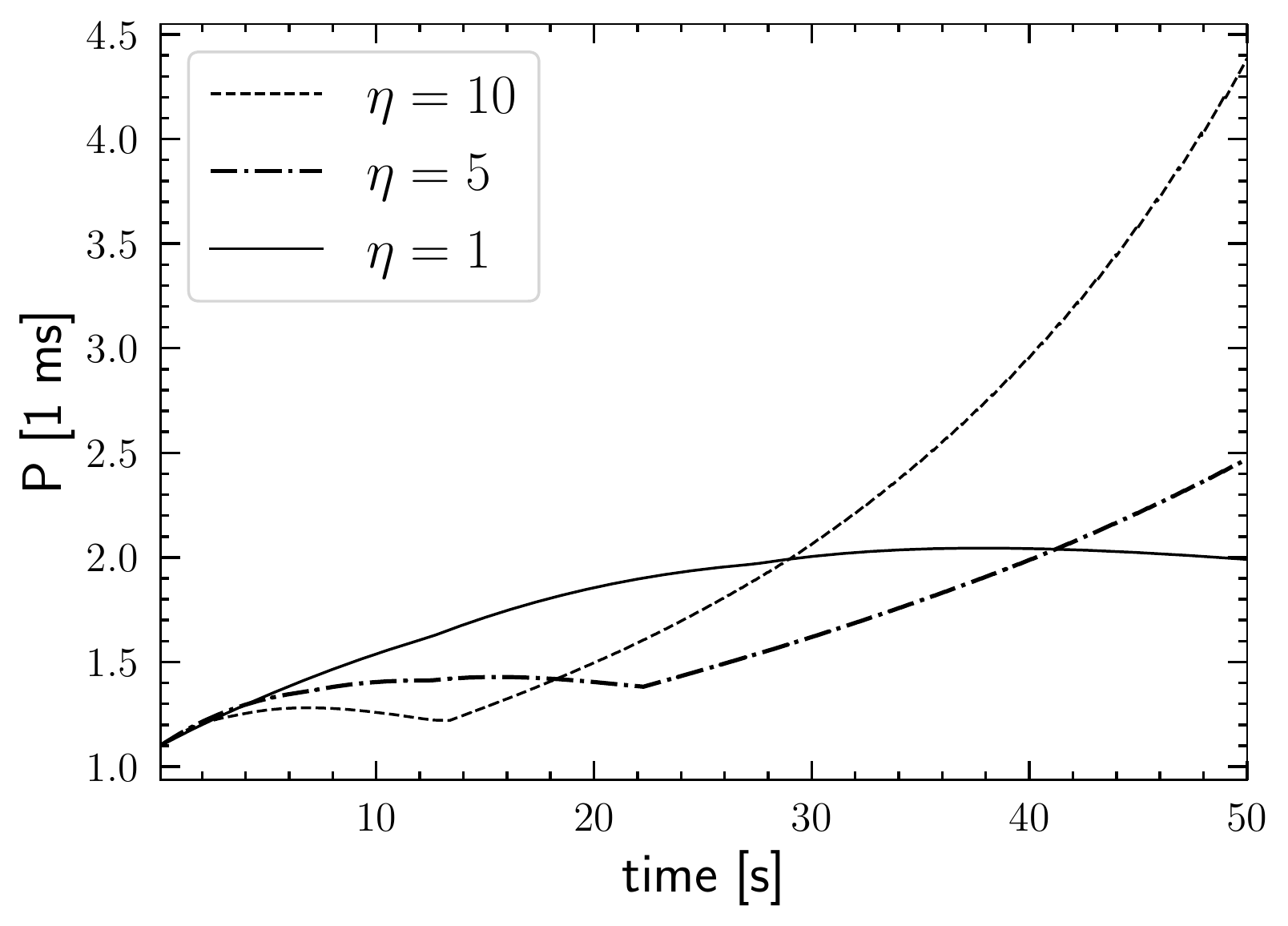}
		\caption{}
		\label{spineta}
	\end{subfigure}%
	\begin{subfigure}{.5\textwidth}
		\centering
		\includegraphics[scale=0.55]{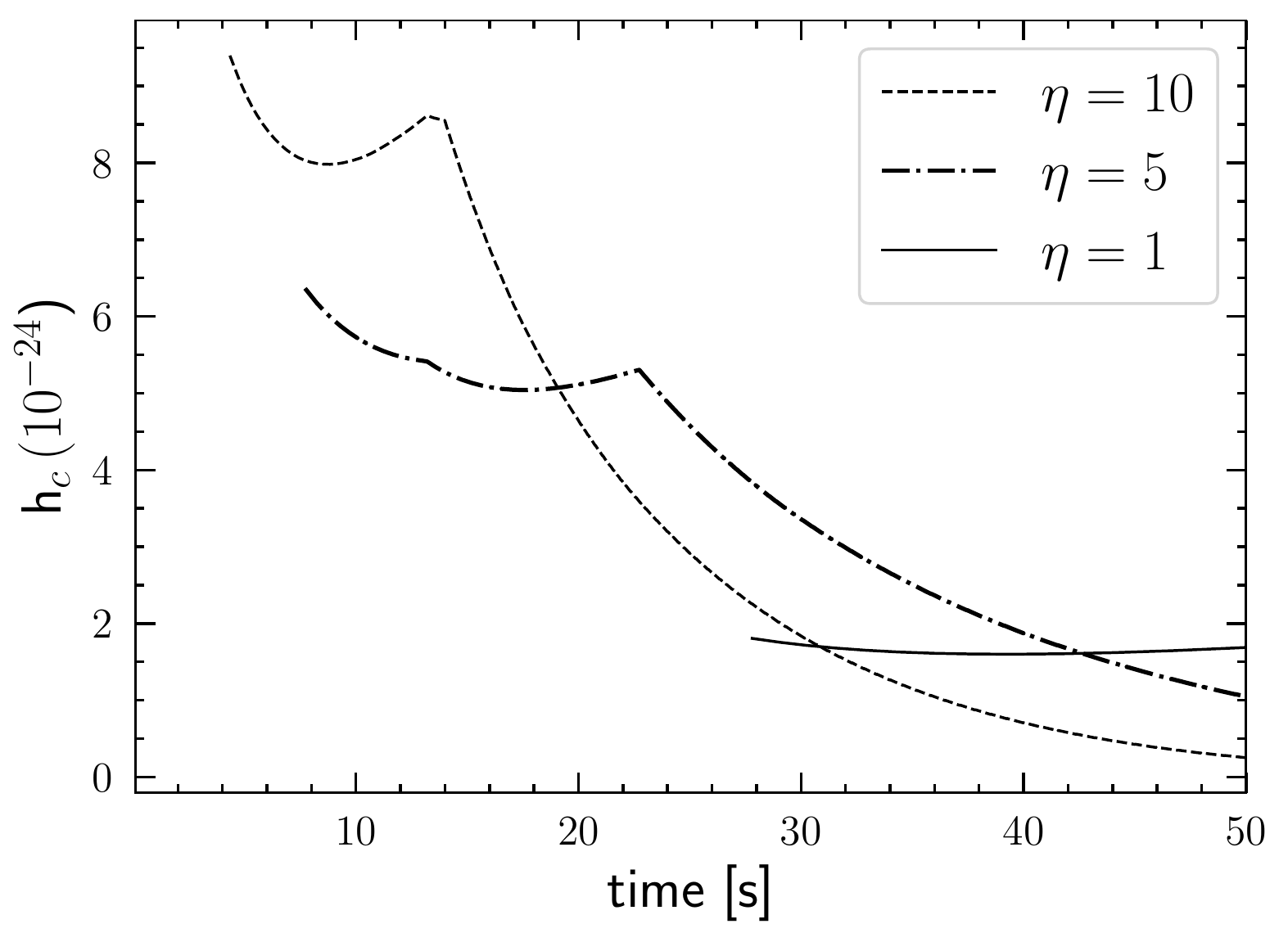}
		\caption{}
		\label{h0eta}
	\end{subfigure}    
	\caption{ (a) Spin evolution and (b) gravitational-wave strains for three different values of $\eta$ for our fiducial magnetar model, formed as a remnant of BNS merger, when the total mass available for accretion is $0.2 \, M_{\odot}$. The initial spin-period is $P_0 = 1.1$ ms. We show results upto 50s when the magnetar survives and not yet collapsed to a BH.}
\end{figure*}

\subsection{Remnant of BNS mergers}
We have assumed so far that following the collapse of the progenitor, there is enough mass available for accretion which increases the total mass until the magnetar collapses to a BH. Let us now look into a scenario in which the magnetar is created in a binary neutron star merger, and  the mass available for accretion, e.g. remnant mass ejected in the merger, is only 0.2 $M_{\odot}$ \citep{Radice2018,Bernuzzi2020, Bernuzzi2020b, Radice2020a}. Although our mass accretion rate $\dot{M} = 10^{-3} \eta t^{1/2} \,\,M_{\odot}s^{-1} $ refers to supernova disk accretion, nonetheless we assume this form for BNS mergers, despite tidal tails formed in a post-merger remnant will accrete back onto the star at a different rate than that through relatively long-term disk accretion. Since in this case, the mass available for accretion is lower compared to the the supernovae disk-accretion, the evolution is  different because the accretion stops soon after the entire mass is accreted and the magnetar spins down. The rate of accretion, in principle, is also not known \footnote{In private communication with Dr. Tim Dietrich.}, which allows us to make a study by varying the parameter $\eta$  to explore a range of values in $\dot{M}$ \citep{Ruiz2017}. The spin evolution for $\eta=10$ is shown in figure \ref{spin2}. Initially, the star loses angular momentum through the neutrino wind and magnetic dipolar radiation until 7 $s$ after which the accretion torque makes it spin faster. In about 14 $s$, the total mass in the disc gets accreted and the torque due to accretion becomes zero. At this point, there is no positive torque on the magnetar and the star continues to spin down. The GW torque is weak compared to the other torques and hence does not influence the overall spin-period, nevertheless the emission is expected to continue, as the mountain is likely to be stable on the dynamical timescales we are studying here, although it is likely to be disrupted on longer timescales as the buried magnetic field re-emerges \citep{Vigelius1, Vigelius2}. We show results upto 50s, although the star survives for longer period of time. The evolution of the spin and GW signal are also shown in figures \ref{spineta} and \ref{h0eta} respectively for $\eta=10$ (dashed line) and compared with two other values, $\eta=5$ (dashed-dotted line) and $\eta=1$ (solid line). We conclude that the spin is mostly dominated by the dipole radiation and the duration of accretion is prolonged on lowering $\eta$ as expected, while GW strain decreases when lowering $\eta$. The shape of the signal can be understood by how the star spins, as an increase in the spin period (decrease in frequency) causes the GW strain to decrease and vice-versa. This signal also remains weaker when compared to the previous case of core-collapsed supernova in which the magnetar spins-up due to accretion.\\

Although we have assumed an upper limit of the mass available for accretion by a remnant of BNS merger, it is possible that the mass available is few orders of magnitude lower (i.e.$0.001-0.01 \, M_{\odot}$). In this case, assuming the same $\dot{M}$, the magnetar would survive longer until its spin frequency decreases significantly due to dipole radiation and it can't hold upto its own mass before collapsing to a BH.

\vspace{0.4 cm}
\section{Summary and Conclusions \label{sec:discussion}}
In this paper we have studied the evolution and gravitational wave emission of a newly born millisecond magnetar having a high accretion rate of $\mathcal{O}(10^{-2}) M_{\odot} s^{-1}$ due to fall back accretion. We assume that accreted matter is confined at the poles and creates a `mountain' that leads to a time varying quadrupole and GW emission \cite{2019PhRvD.100l3014Z}.
If the magnetar is formed due to the collapse of a massive star, depending on its initial mass and rotation rate, survives for a timescale of order $t\approx 50 \,s$ before collapsing to a BH. If, on the other hand, it is formed after the merger of a binary NS system, the magnetar will survive much longer (although as it spins down it can eventually collapse to a black hole \citealt{Ravi14}), but its GW emission will be weaker.

The magnetar experiences different torques shortly after birth. In particular we include for the first time the angular momentum carried away by magnetic dipole radiation, and find it to be the main spin-down mechanism at early times, while neutrino-wind and GWs do not affect the spin period significantly, in comparison.
We also include the effect of magnetic field burial, and of the evolution of the inclination angle between the field and rotation axis. We find also that the magnetic moment axis becomes orthogonal to the rotation axis immediately after the star is born. Generally a long-lived surviving magnetar is expected to have a very small inclination angle \citep{2020MNRAS.494.4838L}. Note however that this can be due to secular evolution of the inclination angle on much longer timescales, and the millisecond magnetars we examine in this paper are unlikely to be the progenitors of the galactic magnetars, as in most cases our systems collapse to a BH.
Finally we use precises relativistic and rapidly rotating models of the magnetar, obtained with the numerical code RNS \citep{RNS}  to calculate the evolution of the gravitational mass, radius and moment of inertia of the star as it accretes matter from the surrounding disc.

Overall, we find that if the magnetar is formed after the collapse of a massive star, there is enough mass in the torus that the accretion torque dominates, spinning up the star. As matter is accreted The GW strain increases with time (mainly due to the increasing frequency of the star), until the star exceeds the maximum mass for its EoS, and collapses to a BH. We expect a ``burst'' signal with an estimated gravitational wave strain $h_c \sim 10^{-24}-10^{-23}$ for objects at a distance of $1\,Mpc$, making them potential targets for the third generation detectors such as the Einstein Telescope and the Cosmic Explorer, although they may be detected by Advanced LIGO and Virgo at design sensitivity should they occur at a distance of a few hundred $kpc$ \citep{LongBurst}. Sensitivity curves to the characteristic strain for these detectors are shown in figure \ref{SNR}. We have further investigated the scenario in which the magnetar is formed as the remnant of a BNS merger. In this case the mass available for accretion is likely to not exceed $0.2 \, M_{\odot}$ \citep{Bernuzzi2020}. We have shown that in this case the combined effects of different processes slows down the spin period also causing the GW strain to decrease with time. In this case the star does not collapse immediately, and is likely to survive for $10^2-10^3$ seconds before it collapses to a BH, if its initial mass exceeds the maximum mass for a non-rotating star \citep{Ravi14}, leading to a signal that may be visible after the merger by next generation detectors \citep{merger}. \cite{Sarin2018PRD} also derived waveform model for millisecond magnetars and showed that the X-ray afterglow can be used to improve search sensitivity by up to $50\%$ and derived horizon distances. The birth-rate of millisecond magnetars associated with super-luminous supernovae is $40 \,\rm Gpc^{-3}yr^{-1}$ while those with LGRBs is $140 \,\rm Gpc^{-3}yr^{-1}$ \citep{birthrate}. BNS merger rates lie between $110-3840 \,\rm Gpc^{-3}yr^{-1}$ \citep{LIGOScientific:2018mvr}. An estimate that $10 \%$ of BNS mergers give birth to millisecond magnetars and the given uncertainties in the formation channels, an average rate of $10-100 \,\rm Gpc^{-3}yr^{-1}$ was reported by \cite{birthrate}. Considering a volume of $1 \rm Mpc^{-3}$, the event rate for this type of GW radiation is less compared to binary neutron star or binary black hole mergers. Finally, we reiterate that the results presented in this paper are model dependent and carry with them many uncertainties. More work is thus needed to explore the different model assumptions that drive the evolution of the system, and produce more robust numerical results.

\section{Acknowledgement}
AS is supported by the OPUS grant from the National Science Centre, Poland (NCN), number 2018/29/B/ST9/02013. We thank the LIGO-Virgo-Kagra collaboration for providing data on the sensitivities for different detectors through the public DCC page \url{https://dcc.ligo.org/LIGO-T1500293-v11/public}. AS and BH particularly thanks Andrew Melatos for his useful and valuable comments on the manuscript. AS further thanks Tim Dietrich for pointing out accretion rates from numerical relativity simulations and group members Giovanni Camelio, Marco Antonelli, and Lorenzo Gavassino for discussions at various stages during this work. BH and AS also thanks Albino Perego for clarifying issues on the neutrino luminosity. This article has been assigned LIGO document number P2000411.

\section{Data Availability}
The data underlying this article will be shared on reasonable request to the corresponding author.

\bibliographystyle{mnras}
\bibliography{references}

\begin{thebibliography}{}
\makeatletter
\relax
\def\mn@urlcharsother{\let\do\@makeother \do\$\do\&\do\#\do\^\do\_\do\%\do\~}
\def\mn@doi{\begingroup\mn@urlcharsother \@ifnextchar [ {\mn@doi@}
  {\mn@doi@[]}}
\def\mn@doi@[#1]#2{\def\@tempa{#1}\ifx\@tempa\@empty \href
  {http://dx.doi.org/#2} {doi:#2}\else \href {http://dx.doi.org/#2} {#1}\fi
  \endgroup}
\def\mn@eprint#1#2{\mn@eprint@#1:#2::\@nil}
\def\mn@eprint@arXiv#1{\href {http://arxiv.org/abs/#1} {{\tt arXiv:#1}}}
\def\mn@eprint@dblp#1{\href {http://dblp.uni-trier.de/rec/bibtex/#1.xml}
  {dblp:#1}}
\def\mn@eprint@#1:#2:#3:#4\@nil{\def\@tempa {#1}\def\@tempb {#2}\def\@tempc
  {#3}\ifx \@tempc \@empty \let \@tempc \@tempb \let \@tempb \@tempa \fi \ifx
  \@tempb \@empty \def\@tempb {arXiv}\fi \@ifundefined
  {mn@eprint@\@tempb}{\@tempb:\@tempc}{\expandafter \expandafter \csname
  mn@eprint@\@tempb\endcsname \expandafter{\@tempc}}}

\bibitem[\protect\citeauthoryear{Abbott et~al.}{Abbott
  et~al.}{2010}]{Collaboration:2009rfa}
Abbott B.,  et~al., 2010, \mn@doi [Astrophys. J.]
  {10.1088/0004-637X/713/1/671}, 713, 671

\bibitem[\protect\citeauthoryear{Abbott et~al.}{Abbott
  et~al.}{2016}]{Abbott:2016blz}
Abbott B.,  et~al., 2016, \mn@doi [Phys. Rev. Lett.]
  {10.1103/PhysRevLett.116.061102}, 116, 061102

\bibitem[\protect\citeauthoryear{{Abbott} et~al.}{{Abbott}
  et~al.}{2017}]{merger}
{Abbott} B.~P.,  et~al., 2017, \mn@doi [\apjl] {10.3847/2041-8213/aa9a35},
  \href {https://ui.adsabs.harvard.edu/abs/2017ApJ...851L..16A} {851, L16}

\bibitem[\protect\citeauthoryear{Abbott et~al.}{Abbott
  et~al.}{2018}]{Abbott:2018exr}
Abbott B.,  et~al., 2018, \mn@doi [Phys. Rev. Lett.]
  {10.1103/PhysRevLett.121.161101}, 121, 161101

\bibitem[\protect\citeauthoryear{Abbott et~al.}{Abbott
  et~al.}{2019a}]{LIGOScientific:2018mvr}
Abbott B.,  et~al., 2019a, \mn@doi [Phys. Rev. X] {10.1103/PhysRevX.9.031040},
  9, 031040

\bibitem[\protect\citeauthoryear{Abbott et~al.}{Abbott
  et~al.}{2019b}]{Abbott:2019heg}
Abbott B.,  et~al., 2019b, \mn@doi [Phys. Rev. D] {10.1103/PhysRevD.99.104033},
  99, 104033

\bibitem[\protect\citeauthoryear{{Abbott} et~al.}{{Abbott}
  et~al.}{2019c}]{LongBurst}
{Abbott} B.~P.,  et~al., 2019c, \mn@doi [\prd] {10.1103/PhysRevD.99.104033},
  \href {https://ui.adsabs.harvard.edu/abs/2019PhRvD..99j4033A} {99, 104033}

\bibitem[\protect\citeauthoryear{Abbott et~al.}{Abbott
  et~al.}{2019d}]{Pisarski:2019vxw}
Abbott B.,  et~al., 2019d, \mn@doi [Phys. Rev. D]
  {10.1103/PhysRevD.100.024004}, 100, 024004

\bibitem[\protect\citeauthoryear{Abbott et~al.}{Abbott
  et~al.}{2019e}]{Abbott:2019prv}
Abbott B.,  et~al., 2019e, \mn@doi [Phys. Rev. D]
  {10.1103/PhysRevD.100.024017}, 100, 024017

\bibitem[\protect\citeauthoryear{Abbott et~al.}{Abbott
  et~al.}{2019f}]{Authors:2019fue}
Abbott B.,  et~al., 2019f, \mn@doi [Astrophys. J.] {10.3847/1538-4357/ab4b48},
  886, 75

\bibitem[\protect\citeauthoryear{Abbott et~al.}{Abbott
  et~al.}{2020a}]{LIGOScientific:2019hgc}
Abbott B.~P.,  et~al., 2020a, \mn@doi [Class. Quant. Grav.]
  {10.1088/1361-6382/ab685e}, 37, 055002

\bibitem[\protect\citeauthoryear{Abbott et~al.}{Abbott
  et~al.}{2020b}]{Abbott:2020uma}
Abbott B.,  et~al., 2020b, \mn@doi [Astrophys. J. Lett.]
  {10.3847/2041-8213/ab75f5}, 892, L3

\bibitem[\protect\citeauthoryear{Abbott et~al.}{Abbott
  et~al.}{2020c}]{Abbott:2020khf}
Abbott R.,  et~al., 2020c, \mn@doi [Astrophys. J.] {10.3847/2041-8213/ab960f},
  896, L44

\bibitem[\protect\citeauthoryear{{Alford} \& {Schwenzer}}{{Alford} \&
  {Schwenzer}}{2014}]{2014ApJ...781...26A}
{Alford} M.~G.,  {Schwenzer} K.,  2014, \mn@doi [\apj]
  {10.1088/0004-637X/781/1/26}, \href
  {https://ui.adsabs.harvard.edu/abs/2014ApJ...781...26A} {781, 26}

\bibitem[\protect\citeauthoryear{{Andersson} \& {Kokkotas}}{{Andersson} \&
  {Kokkotas}}{2001}]{2001IJMPD..10..381A}
{Andersson} N.,  {Kokkotas} K.~D.,  2001, \mn@doi [International Journal of
  Modern Physics D] {10.1142/S0218271801001062}, \href
  {https://ui.adsabs.harvard.edu/abs/2001IJMPD..10..381A} {10, 381}

\bibitem[\protect\citeauthoryear{{Archibald} et~al.,}{{Archibald}
  et~al.}{2016}]{Archibald2016}
{Archibald} R.~F.,  et~al., 2016, \mn@doi [\apjl]
  {10.3847/2041-8205/819/1/L16}, \href
  {https://ui.adsabs.harvard.edu/abs/2016ApJ...819L..16A} {819, L16}

\bibitem[\protect\citeauthoryear{{Basko} \& {Sunyaev}}{{Basko} \&
  {Sunyaev}}{1976}]{1976MNRAS.175..395B}
{Basko} M.~M.,  {Sunyaev} R.~A.,  1976, \mn@doi [\mnras]
  {10.1093/mnras/175.2.395}, \href
  {https://ui.adsabs.harvard.edu/abs/1976MNRAS.175..395B} {175, 395}

\bibitem[\protect\citeauthoryear{Baumgarte \& Shapiro}{Baumgarte \&
  Shapiro}{1998}]{Baumgarte_1998}
Baumgarte T.~W.,  Shapiro S.~L.,  1998, \mn@doi [The Astrophysical Journal]
  {10.1086/306067}, 504, 431

\bibitem[\protect\citeauthoryear{{Bernuzzi}}{{Bernuzzi}}{2020}]{Bernuzzi2020}
{Bernuzzi} S.,  2020, arXiv e-prints, \href
  {https://ui.adsabs.harvard.edu/abs/2020arXiv200406419B} {p. arXiv:2004.06419}

\bibitem[\protect\citeauthoryear{{Bernuzzi} et~al.,}{{Bernuzzi}
  et~al.}{2020}]{Bernuzzi2020b}
{Bernuzzi} S.,  et~al., 2020, \mn@doi [\mnras] {10.1093/mnras/staa1860}, \href
  {https://ui.adsabs.harvard.edu/abs/2020MNRAS.tmp.1981B} {}

\bibitem[\protect\citeauthoryear{{Bondarescu}, {Teukolsky}  \&
  {Wasserman}}{{Bondarescu} et~al.}{2007}]{2007PhRvD..76f4019B}
{Bondarescu} R.,  {Teukolsky} S.~A.,   {Wasserman} I.,  2007, \mn@doi [\prd]
  {10.1103/PhysRevD.76.064019}, \href
  {https://ui.adsabs.harvard.edu/abs/2007PhRvD..76f4019B} {76, 064019}

\bibitem[\protect\citeauthoryear{{Breu} \& {Rezzolla}}{{Breu} \&
  {Rezzolla}}{2016}]{Breu16}
{Breu} C.,  {Rezzolla} L.,  2016, \mn@doi [\mnras] {10.1093/mnras/stw575},
  \href {https://ui.adsabs.harvard.edu/abs/2016MNRAS.459..646B} {459, 646}

\bibitem[\protect\citeauthoryear{{Brown} \& {Bildsten}}{{Brown} \&
  {Bildsten}}{1998}]{Brown98}
{Brown} E.~F.,  {Bildsten} L.,  1998, \mn@doi [\apj] {10.1086/305419}, \href
  {https://ui.adsabs.harvard.edu/abs/1998ApJ...496..915B} {496, 915}

\bibitem[\protect\citeauthoryear{{Bucciantini}, {Quataert}, {Arons}, {Metzger}
  \& {Thompson}}{{Bucciantini} et~al.}{2008}]{2008MNRAS.383L..25B}
{Bucciantini} N.,  {Quataert} E.,  {Arons} J.,  {Metzger} B.~D.,   {Thompson}
  T.~A.,  2008, \mn@doi [\mnras] {10.1111/j.1745-3933.2007.00403.x}, \href
  {https://ui.adsabs.harvard.edu/abs/2008MNRAS.383L..25B} {383, L25}

\bibitem[\protect\citeauthoryear{{Bucciantini}, {Quataert}, {Metzger},
  {Thompson}, {Arons}  \& {Del Zanna}}{{Bucciantini}
  et~al.}{2009}]{2009MNRAS.396.2038B}
{Bucciantini} N.,  {Quataert} E.,  {Metzger} B.~D.,  {Thompson} T.~A.,  {Arons}
  J.,   {Del Zanna} L.,  2009, \mn@doi [\mnras]
  {10.1111/j.1365-2966.2009.14940.x}, \href
  {https://ui.adsabs.harvard.edu/abs/2009MNRAS.396.2038B} {396, 2038}

\bibitem[\protect\citeauthoryear{Ciolfi, Lander, Manca  \& Rezzolla}{Ciolfi
  et~al.}{2011}]{Ciolfi:2011xa}
Ciolfi R.,  Lander S.~K.,  Manca G.~M.,   Rezzolla L.,  2011, \mn@doi
  [Astrophys. J. Lett.] {10.1088/2041-8205/736/1/L6}, 736, L6

\bibitem[\protect\citeauthoryear{Corsi \& M\'esz\'aros}{Corsi \&
  M\'esz\'aros}{2009}]{Corsi2009}
Corsi A.,  M\'esz\'aros P.,  2009, \mn@doi [Astrophys. J.]
  {10.1088/0004-637X/702/2/1171}, 702, 1171

\bibitem[\protect\citeauthoryear{{Cromartie} et~al.,}{{Cromartie}
  et~al.}{2020}]{Cromartie2020}
{Cromartie} H.~T.,  et~al., 2020, \mn@doi [Nature Astronomy]
  {10.1038/s41550-019-0880-2}, \href
  {https://ui.adsabs.harvard.edu/abs/2020NatAs...4...72C} {4, 72}

\bibitem[\protect\citeauthoryear{Cutler}{Cutler}{2002}]{Cutler:2002nw}
Cutler C.,  2002, \mn@doi [Phys. Rev. D] {10.1103/PhysRevD.66.084025}, 66,
  084025

\bibitem[\protect\citeauthoryear{{Dai} \& {Liu}}{{Dai} \&
  {Liu}}{2012}]{Dai2012}
{Dai} Z.~G.,  {Liu} R.-Y.,  2012, \mn@doi [\apj] {10.1088/0004-637X/759/1/58},
  \href {https://ui.adsabs.harvard.edu/abs/2012ApJ...759...58D} {759, 58}

\bibitem[\protect\citeauthoryear{{Dai} \& {Lu}}{{Dai} \&
  {Lu}}{1998}]{1998PhRvL..81.4301D}
{Dai} Z.~G.,  {Lu} T.,  1998, \mn@doi [\prl] {10.1103/PhysRevLett.81.4301},
  \href {https://ui.adsabs.harvard.edu/abs/1998PhRvL..81.4301D} {81, 4301}

\bibitem[\protect\citeauthoryear{{Dall'Osso} \& {Perna}}{{Dall'Osso} \&
  {Perna}}{2017}]{2017MNRAS.472.2142D}
{Dall'Osso} S.,  {Perna} R.,  2017, \mn@doi [\mnras] {10.1093/mnras/stx2097},
  \href {https://ui.adsabs.harvard.edu/abs/2017MNRAS.472.2142D} {472, 2142}

\bibitem[\protect\citeauthoryear{Dall'Osso, Stella  \& Palomba}{Dall'Osso
  et~al.}{2018}]{DallOsso:2018dos}
Dall'Osso S.,  Stella L.,   Palomba C.,  2018, \mn@doi [Mon. Not. Roy. Astron.
  Soc.] {10.1093/mnras/sty1706}, 480, 1353

\bibitem[\protect\citeauthoryear{{Demorest}, {Pennucci}, {Ransom}, {Roberts}
  \& {Hessels}}{{Demorest} et~al.}{2010}]{2010Natur.467.1081D}
{Demorest} P.~B.,  {Pennucci} T.,  {Ransom} S.~M.,  {Roberts} M.~S.~E.,
  {Hessels} J.~W.~T.,  2010, \mn@doi [\nat] {10.1038/nature09466}, \href
  {https://ui.adsabs.harvard.edu/abs/2010Natur.467.1081D} {467, 1081}

\bibitem[\protect\citeauthoryear{{Duncan} \& {Thompson}}{{Duncan} \&
  {Thompson}}{1992}]{1992ApJ...392L...9D}
{Duncan} R.~C.,  {Thompson} C.,  1992, \mn@doi [\apjl] {10.1086/186413}, \href
  {https://ui.adsabs.harvard.edu/abs/1992ApJ...392L...9D} {392, L9}

\bibitem[\protect\citeauthoryear{{Ek{\textcommabelow s}i}, {Hernquist}  \&
  {Narayan}}{{Ek{\textcommabelow s}i} et~al.}{2005}]{Eksi2005}
{Ek{\textcommabelow s}i} K.~Y.,  {Hernquist} L.,   {Narayan} R.,  2005, \mn@doi
  [\apjl] {10.1086/429915}, \href
  {https://ui.adsabs.harvard.edu/abs/2005ApJ...623L..41E} {623, L41}

\bibitem[\protect\citeauthoryear{{Erkut}, {Ek{\textcommabelow s}i}  \&
  {Alpar}}{{Erkut} et~al.}{2019}]{Erkut2019}
{Erkut} M.~H.,  {Ek{\textcommabelow s}i} K.~Y.,   {Alpar} M.~A.,  2019, \mn@doi
  [\apj] {10.3847/1538-4357/ab04ae}, \href
  {https://ui.adsabs.harvard.edu/abs/2019ApJ...873..105E} {873, 105}

\bibitem[\protect\citeauthoryear{{Giacomazzo} \& {Perna}}{{Giacomazzo} \&
  {Perna}}{2013}]{2013ApJ...771L..26G}
{Giacomazzo} B.,  {Perna} R.,  2013, \mn@doi [\apjl]
  {10.1088/2041-8205/771/2/L26}, \href
  {https://ui.adsabs.harvard.edu/abs/2013ApJ...771L..26G} {771, L26}

\bibitem[\protect\citeauthoryear{{Gittins}, {Andersson}  \& {Jones}}{{Gittins}
  et~al.}{2020}]{Gittins20}
{Gittins} F.,  {Andersson} N.,   {Jones} D.~I.,  2020, arXiv e-prints, \href
  {https://ui.adsabs.harvard.edu/abs/2020arXiv200912794G} {p. arXiv:2009.12794}

\bibitem[\protect\citeauthoryear{Gusakov \& Chugunov}{Gusakov \&
  Chugunov}{2020}]{Gusakov:2020}
Gusakov M.~E.,  Chugunov A.~I.,  2020, \mn@doi [Phys. Rev. Lett.]
  {10.1103/PhysRevLett.124.191101}, 124, 191101

\bibitem[\protect\citeauthoryear{{Haensel} \& {Zdunik}}{{Haensel} \&
  {Zdunik}}{2008}]{Haensel2008}
{Haensel} P.,  {Zdunik} J.~L.,  2008, \mn@doi [\aap]
  {10.1051/0004-6361:20078578}, \href
  {https://ui.adsabs.harvard.edu/abs/2008A&A...480..459H} {480, 459}

\bibitem[\protect\citeauthoryear{{Haensel}, {Potekhin}  \&
  {Yakovlev}}{{Haensel} et~al.}{2007}]{HaenselBook}
{Haensel} P.,  {Potekhin} A.~Y.,   {Yakovlev} D.~G.,  2007, {Neutron Stars 1 :
  Equation of State and Structure}.
 Vol. 326

\bibitem[\protect\citeauthoryear{{Haskell}}{{Haskell}}{2015}]{2015IJMPE..2441007H}
{Haskell} B.,  2015, \mn@doi [International Journal of Modern Physics E]
  {10.1142/S0218301315410074}, \href
  {https://ui.adsabs.harvard.edu/abs/2015IJMPE..2441007H} {24, 1541007}

\bibitem[\protect\citeauthoryear{{Haskell}, {Andersson}  \&
  {Passamonti}}{{Haskell} et~al.}{2009}]{2009MNRAS.397.1464H}
{Haskell} B.,  {Andersson} N.,   {Passamonti} A.,  2009, \mn@doi [\mnras]
  {10.1111/j.1365-2966.2009.14963.x}, \href
  {https://ui.adsabs.harvard.edu/abs/2009MNRAS.397.1464H} {397, 1464}

\bibitem[\protect\citeauthoryear{{Haskell} et~al.,}{{Haskell}
  et~al.}{2015a}]{2015ASSP...40...85H}
{Haskell} B.,  et~al., 2015a, in Gravitational Wave Astrophysics. p.~85
  (\mn@eprint {arXiv} {1407.8254}), \mn@doi{10.1007/978-3-319-10488-1_8}

\bibitem[\protect\citeauthoryear{{Haskell}, {Priymak}, {Patruno}, {Oppenoorth},
  {Melatos}  \& {Lasky}}{{Haskell} et~al.}{2015b}]{2015MNRAS.450.2393H}
{Haskell} B.,  {Priymak} M.,  {Patruno} A.,  {Oppenoorth} M.,  {Melatos} A.,
  {Lasky} P.~D.,  2015b, \mn@doi [\mnras] {10.1093/mnras/stv726}, \href
  {https://ui.adsabs.harvard.edu/abs/2015MNRAS.450.2393H} {450, 2393}

\bibitem[\protect\citeauthoryear{{Jones}}{{Jones}}{1976}]{1976Ap&SS..45..369J}
{Jones} P.~B.,  1976, \mn@doi [\apss] {10.1007/BF00642671}, \href
  {https://ui.adsabs.harvard.edu/abs/1976Ap&SS..45..369J} {45, 369}

\bibitem[\protect\citeauthoryear{{Lander} \& {Jones}}{{Lander} \&
  {Jones}}{2020}]{2020MNRAS.494.4838L}
{Lander} S.~K.,  {Jones} D.~I.,  2020, \mn@doi [\mnras]
  {10.1093/mnras/staa966}, \href
  {https://ui.adsabs.harvard.edu/abs/2020MNRAS.494.4838L} {494, 4838}

\bibitem[\protect\citeauthoryear{Lasky}{Lasky}{2015}]{Lasky:2015uia}
Lasky P.~D.,  2015, \mn@doi [Publ. Astron. Soc. Austral.]
  {10.1017/pasa.2015.35}, 32, e034

\bibitem[\protect\citeauthoryear{{Lasky}, {Haskell}, {Ravi}, {Howell}  \&
  {Coward}}{{Lasky} et~al.}{2014}]{HaskellRavi}
{Lasky} P.~D.,  {Haskell} B.,  {Ravi} V.,  {Howell} E.~J.,   {Coward} D.~M.,
  2014, \mn@doi [\prd] {10.1103/PhysRevD.89.047302}, \href
  {https://ui.adsabs.harvard.edu/abs/2014PhRvD..89d7302L} {89, 047302}

\bibitem[\protect\citeauthoryear{{Lasky}, {Leris}, {Rowlinson}  \&
  {Glampedakis}}{{Lasky} et~al.}{2017}]{Lasky2017}
{Lasky} P.~D.,  {Leris} C.,  {Rowlinson} A.,   {Glampedakis} K.,  2017, \mn@doi
  [\apjl] {10.3847/2041-8213/aa79a7}, \href
  {https://ui.adsabs.harvard.edu/abs/2017ApJ...843L...1L} {843, L1}

\bibitem[\protect\citeauthoryear{{Lattimer} \& {Schutz}}{{Lattimer} \&
  {Schutz}}{2005}]{LattimerSchutz}
{Lattimer} J.~M.,  {Schutz} B.~F.,  2005, \mn@doi [\apj] {10.1086/431543},
  \href {https://ui.adsabs.harvard.edu/abs/2005ApJ...629..979L} {629, 979}

\bibitem[\protect\citeauthoryear{{MacFadyen}, {Woosley}  \&
  {Heger}}{{MacFadyen} et~al.}{2001}]{MacFayden2001}
{MacFadyen} A.~I.,  {Woosley} S.~E.,   {Heger} A.,  2001, \mn@doi [\apj]
  {10.1086/319698}, \href
  {https://ui.adsabs.harvard.edu/abs/2001ApJ...550..410M} {550, 410}

\bibitem[\protect\citeauthoryear{{Melatos}}{{Melatos}}{1997}]{Melatos97}
{Melatos} A.,  1997, \mn@doi [\mnras] {10.1093/mnras/288.4.1049}, \href
  {https://ui.adsabs.harvard.edu/abs/1997MNRAS.288.1049M} {288, 1049}

\bibitem[\protect\citeauthoryear{{Melatos} \& {Payne}}{{Melatos} \&
  {Payne}}{2005}]{2005ApJ...623.1044M}
{Melatos} A.,  {Payne} D.~J.~B.,  2005, \mn@doi [\apj] {10.1086/428600}, \href
  {https://ui.adsabs.harvard.edu/abs/2005ApJ...623.1044M} {623, 1044}

\bibitem[\protect\citeauthoryear{{Melatos} \& {Priymak}}{{Melatos} \&
  {Priymak}}{2014}]{Melatos14}
{Melatos} A.,  {Priymak} M.,  2014, \mn@doi [\apj]
  {10.1088/0004-637X/794/2/170}, \href
  {https://ui.adsabs.harvard.edu/abs/2014ApJ...794..170M} {794, 170}

\bibitem[\protect\citeauthoryear{{M{\'e}sz{\'a}ros}}{{M{\'e}sz{\'a}ros}}{2006}]{MezRev}
{M{\'e}sz{\'a}ros} P.,  2006, \mn@doi [Reports on Progress in Physics]
  {10.1088/0034-4885/69/8/R01}, \href
  {https://ui.adsabs.harvard.edu/abs/2006RPPh...69.2259M} {69, 2259}

\bibitem[\protect\citeauthoryear{{Metzger}, {Giannios}, {Thompson},
  {Bucciantini}  \& {Quataert}}{{Metzger} et~al.}{2011}]{Mertzger2011}
{Metzger} B.~D.,  {Giannios} D.,  {Thompson} T.~A.,  {Bucciantini} N.,
  {Quataert} E.,  2011, \mn@doi [\mnras] {10.1111/j.1365-2966.2011.18280.x},
  \href {https://ui.adsabs.harvard.edu/abs/2011MNRAS.413.2031M} {413, 2031}

\bibitem[\protect\citeauthoryear{{Murase} \& {Bartos}}{{Murase} \&
  {Bartos}}{2019}]{MuraseRev}
{Murase} K.,  {Bartos} I.,  2019, \mn@doi [Annual Review of Nuclear and
  Particle Science] {10.1146/annurev-nucl-101918-023510}, \href
  {https://ui.adsabs.harvard.edu/abs/2019ARNPS..69..477M} {69, 477}

\bibitem[\protect\citeauthoryear{{Nicholl}, {Williams}, {Berger}, {Villar},
  {Alexander}, {Eftekhari}  \& {Metzger}}{{Nicholl} et~al.}{2017}]{birthrate}
{Nicholl} M.,  {Williams} P.~K.~G.,  {Berger} E.,  {Villar} V.~A.,  {Alexander}
  K.~D.,  {Eftekhari} T.,   {Metzger} B.~D.,  2017, \mn@doi [\apj]
  {10.3847/1538-4357/aa794d}, \href
  {https://ui.adsabs.harvard.edu/abs/2017ApJ...843...84N} {843, 84}

\bibitem[\protect\citeauthoryear{{Ott}, {Burrows}, {Thompson}, {Livne}  \&
  {Walder}}{{Ott} et~al.}{2006}]{Ott2006}
{Ott} C.~D.,  {Burrows} A.,  {Thompson} T.~A.,  {Livne} E.,   {Walder} R.,
  2006, \mn@doi [\apjs] {10.1086/500832}, \href
  {https://ui.adsabs.harvard.edu/abs/2006ApJS..164..130O} {164, 130}

\bibitem[\protect\citeauthoryear{{Owen}, {Lindblom}, {Cutler}, {Schutz},
  {Vecchio}  \& {Andersson}}{{Owen} et~al.}{1998}]{Owen1998}
{Owen} B.~J.,  {Lindblom} L.,  {Cutler} C.,  {Schutz} B.~F.,  {Vecchio} A.,
  {Andersson} N.,  1998, \mn@doi [\prd] {10.1103/PhysRevD.58.084020}, \href
  {https://ui.adsabs.harvard.edu/abs/1998PhRvD..58h4020O} {58, 084020}

\bibitem[\protect\citeauthoryear{{Payne} \& {Melatos}}{{Payne} \&
  {Melatos}}{2004}]{Payne2004}
{Payne} D.~J.~B.,  {Melatos} A.,  2004, \mn@doi [\mnras]
  {10.1111/j.1365-2966.2004.07798.x}, \href
  {https://ui.adsabs.harvard.edu/abs/2004MNRAS.351..569P} {351, 569}

\bibitem[\protect\citeauthoryear{{Piro} \& {Ott}}{{Piro} \& {Ott}}{2011}]{Piro}
{Piro} A.~L.,  {Ott} C.~D.,  2011, \mn@doi [\apj]
  {10.1088/0004-637X/736/2/108}, \href
  {https://ui.adsabs.harvard.edu/abs/2011ApJ...736..108P} {736, 108}

\bibitem[\protect\citeauthoryear{{Piro} \& {Thrane}}{{Piro} \&
  {Thrane}}{2012}]{Piro12}
{Piro} A.~L.,  {Thrane} E.,  2012, \mn@doi [\apj] {10.1088/0004-637X/761/1/63},
  \href {https://ui.adsabs.harvard.edu/abs/2012ApJ...761...63P} {761, 63}

\bibitem[\protect\citeauthoryear{{Pons}, {Reddy}, {Prakash}, {Lattimer}  \&
  {Miralles}}{{Pons} et~al.}{1999}]{Pons1999}
{Pons} J.~A.,  {Reddy} S.,  {Prakash} M.,  {Lattimer} J.~M.,   {Miralles}
  J.~A.,  1999, \mn@doi [\apj] {10.1086/306889}, \href
  {https://ui.adsabs.harvard.edu/abs/1999ApJ...513..780P} {513, 780}

\bibitem[\protect\citeauthoryear{{Radice}, {Perego}, {Hotokezaka}, {Fromm},
  {Bernuzzi}  \& {Roberts}}{{Radice} et~al.}{2018}]{Radice2018}
{Radice} D.,  {Perego} A.,  {Hotokezaka} K.,  {Fromm} S.~A.,  {Bernuzzi} S.,
  {Roberts} L.~F.,  2018, \mn@doi [\apj] {10.3847/1538-4357/aaf054}, \href
  {https://ui.adsabs.harvard.edu/abs/2018ApJ...869..130R} {869, 130}

\bibitem[\protect\citeauthoryear{{Radice}, {Bernuzzi}  \& {Perego}}{{Radice}
  et~al.}{2020}]{Radice2020a}
{Radice} D.,  {Bernuzzi} S.,   {Perego} A.,  2020, \mn@doi [Annual Review of
  Nuclear and Particle Science] {10.1146/annurev-nucl-013120-114541}, \href
  {https://ui.adsabs.harvard.edu/abs/2020ARNPS..7013120R} {70, annurev}

\bibitem[\protect\citeauthoryear{{Ravi} \& {Lasky}}{{Ravi} \&
  {Lasky}}{2014}]{Ravi14}
{Ravi} V.,  {Lasky} P.~D.,  2014, \mn@doi [\mnras] {10.1093/mnras/stu720},
  \href {https://ui.adsabs.harvard.edu/abs/2014MNRAS.441.2433R} {441, 2433}

\bibitem[\protect\citeauthoryear{{Rowlinson} et~al.,}{{Rowlinson}
  et~al.}{2010}]{Rowlinson2010}
{Rowlinson} A.,  et~al., 2010, \mn@doi [\mnras]
  {10.1111/j.1365-2966.2010.17354.x}, \href
  {https://ui.adsabs.harvard.edu/abs/2010MNRAS.409..531R} {409, 531}

\bibitem[\protect\citeauthoryear{{Rowlinson}, {O'Brien}, {Metzger}, {Tanvir}
  \& {Levan}}{{Rowlinson} et~al.}{2013}]{Rowlinson2013}
{Rowlinson} A.,  {O'Brien} P.~T.,  {Metzger} B.~D.,  {Tanvir} N.~R.,   {Levan}
  A.~J.,  2013, \mn@doi [\mnras] {10.1093/mnras/sts683}, \href
  {https://ui.adsabs.harvard.edu/abs/2013MNRAS.430.1061R} {430, 1061}

\bibitem[\protect\citeauthoryear{{Ruiz} \& {Shapiro}}{{Ruiz} \&
  {Shapiro}}{2017}]{Ruiz2017}
{Ruiz} M.,  {Shapiro} S.~L.,  2017, \mn@doi [\prd]
  {10.1103/PhysRevD.96.084063}, \href
  {https://ui.adsabs.harvard.edu/abs/2017PhRvD..96h4063R} {96, 084063}

\bibitem[\protect\citeauthoryear{{Sarin}, {Lasky}, {Sammut}  \&
  {Ashton}}{{Sarin} et~al.}{2018}]{Sarin2018PRD}
{Sarin} N.,  {Lasky} P.~D.,  {Sammut} L.,   {Ashton} G.,  2018, \mn@doi [\prd]
  {10.1103/PhysRevD.98.043011}, \href
  {https://ui.adsabs.harvard.edu/abs/2018PhRvD..98d3011S} {98, 043011}

\bibitem[\protect\citeauthoryear{{Sarin}, {Lasky}  \& {Ashton}}{{Sarin}
  et~al.}{2019}]{Sarin2019}
{Sarin} N.,  {Lasky} P.~D.,   {Ashton} G.,  2019, \mn@doi [\apj]
  {10.3847/1538-4357/aaf9a0}, \href
  {https://ui.adsabs.harvard.edu/abs/2019ApJ...872..114S} {872, 114}

\bibitem[\protect\citeauthoryear{{Sarin}, {Lasky}  \& {Ashton}}{{Sarin}
  et~al.}{2020}]{Sarin2020}
{Sarin} N.,  {Lasky} P.~D.,   {Ashton} G.,  2020, \mn@doi [\prd]
  {10.1103/PhysRevD.101.063021}, \href
  {https://ui.adsabs.harvard.edu/abs/2020PhRvD.101f3021S} {101, 063021}

\bibitem[\protect\citeauthoryear{{Schwab}, {Quataert}  \& {Bildsten}}{{Schwab}
  et~al.}{2015}]{2015MNRAS.453.1910S}
{Schwab} J.,  {Quataert} E.,   {Bildsten} L.,  2015, \mn@doi [\mnras]
  {10.1093/mnras/stv1804}, \href
  {https://ui.adsabs.harvard.edu/abs/2015MNRAS.453.1910S} {453, 1910}

\bibitem[\protect\citeauthoryear{{Singh}, {Haskell}, {Mukherjee}  \&
  {Bulik}}{{Singh} et~al.}{2020}]{Singh20}
{Singh} N.,  {Haskell} B.,  {Mukherjee} D.,   {Bulik} T.,  2020, \mn@doi
  [\mnras] {10.1093/mnras/staa442}, \href
  {https://ui.adsabs.harvard.edu/abs/2020MNRAS.493.3866S} {493, 3866}

\bibitem[\protect\citeauthoryear{Spitkovsky}{Spitkovsky}{2006}]{Spitkovsky:2006np}
Spitkovsky A.,  2006, \mn@doi [Astrophys. J. Lett.] {10.1086/507518}, 648, L51

\bibitem[\protect\citeauthoryear{{Stergioulas} \& {Friedman}}{{Stergioulas} \&
  {Friedman}}{1995}]{RNS}
{Stergioulas} N.,  {Friedman} J.~L.,  1995, \mn@doi [\apj] {10.1086/175605},
  \href {https://ui.adsabs.harvard.edu/abs/1995ApJ...444..306S} {444, 306}

\bibitem[\protect\citeauthoryear{{Strang} \& {Melatos}}{{Strang} \&
  {Melatos}}{2019}]{Strang2019}
{Strang} L.~C.,  {Melatos} A.,  2019, \mn@doi [\mnras] {10.1093/mnras/stz1648},
  \href {https://ui.adsabs.harvard.edu/abs/2019MNRAS.487.5010S} {487, 5010}

\bibitem[\protect\citeauthoryear{{Tauris}, {Sanyal}, {Yoon}  \&
  {Langer}}{{Tauris} et~al.}{2013}]{2013A&A...558A..39T}
{Tauris} T.~M.,  {Sanyal} D.,  {Yoon} S.~C.,   {Langer} N.,  2013, \mn@doi
  [\aap] {10.1051/0004-6361/201321662}, \href
  {https://ui.adsabs.harvard.edu/abs/2013A&A...558A..39T} {558, A39}

\bibitem[\protect\citeauthoryear{{Thompson} \& {Duncan}}{{Thompson} \&
  {Duncan}}{1993}]{1993ApJ...408..194T}
{Thompson} C.,  {Duncan} R.~C.,  1993, \mn@doi [\apj] {10.1086/172580}, \href
  {https://ui.adsabs.harvard.edu/abs/1993ApJ...408..194T} {408, 194}

\bibitem[\protect\citeauthoryear{Thompson, Chang  \& Quataert}{Thompson
  et~al.}{2004}]{Thompson:2004wi}
Thompson T.~A.,  Chang P.,   Quataert E.,  2004, \mn@doi [Astrophys. J.]
  {10.1086/421969}, 611, 380

\bibitem[\protect\citeauthoryear{{Ushomirsky}, {Cutler}  \&
  {Bildsten}}{{Ushomirsky} et~al.}{2000}]{Ushomirsky2000}
{Ushomirsky} G.,  {Cutler} C.,   {Bildsten} L.,  2000, \mn@doi [\mnras]
  {10.1046/j.1365-8711.2000.03938.x}, \href
  {https://ui.adsabs.harvard.edu/abs/2000MNRAS.319..902U} {319, 902}

\bibitem[\protect\citeauthoryear{Usov}{Usov}{1992}]{Usov:1992zd}
Usov V.~V.,  1992, \mn@doi [Nature] {10.1038/357472a0}, 357, 472

\bibitem[\protect\citeauthoryear{{Vigelius} \& {Melatos}}{{Vigelius} \&
  {Melatos}}{2009a}]{Vigelius2}
{Vigelius} M.,  {Melatos} A.,  2009a, \mn@doi [\mnras]
  {10.1111/j.1365-2966.2009.14690.x}, \href
  {https://ui.adsabs.harvard.edu/abs/2009MNRAS.395.1972V} {395, 1972}

\bibitem[\protect\citeauthoryear{{Vigelius} \& {Melatos}}{{Vigelius} \&
  {Melatos}}{2009b}]{Vigelius1}
{Vigelius} M.,  {Melatos} A.,  2009b, \mn@doi [\mnras]
  {10.1111/j.1365-2966.2009.14698.x}, \href
  {https://ui.adsabs.harvard.edu/abs/2009MNRAS.395.1985V} {395, 1985}

\bibitem[\protect\citeauthoryear{{Wheeler}, {Yi}, {H{\"o}flich}  \&
  {Wang}}{{Wheeler} et~al.}{2000}]{2000ApJ...537..810W}
{Wheeler} J.~C.,  {Yi} I.,  {H{\"o}flich} P.,   {Wang} L.,  2000, \mn@doi
  [\apj] {10.1086/309055}, \href
  {https://ui.adsabs.harvard.edu/abs/2000ApJ...537..810W} {537, 810}

\bibitem[\protect\citeauthoryear{{Zhang} \& {Dai}}{{Zhang} \&
  {Dai}}{2008}]{Zhang2008}
{Zhang} D.,  {Dai} Z.~G.,  2008, \mn@doi [\apj] {10.1086/589820}, \href
  {https://ui.adsabs.harvard.edu/abs/2008ApJ...683..329Z} {683, 329}

\bibitem[\protect\citeauthoryear{{Zhong}, {Dai}  \& {Li}}{{Zhong}
  et~al.}{2019}]{2019PhRvD.100l3014Z}
{Zhong} S.-Q.,  {Dai} Z.-G.,   {Li} X.-D.,  2019, \mn@doi [\prd]
  {10.1103/PhysRevD.100.123014}, \href
  {https://ui.adsabs.harvard.edu/abs/2019PhRvD.100l3014Z} {100, 123014}

\bibitem[\protect\citeauthoryear{{{\c{C}}{\i}k{\i}nto{\u{g}}lu},
  {{\c{S}}a{\textcommabelow s}maz Mu{\textcommabelow s}}  \&
  {Ek{\textcommabelow s}i}}{{{\c{C}}{\i}k{\i}nto{\u{g}}lu}
  et~al.}{2020}]{Samaz2020}
{{\c{C}}{\i}k{\i}nto{\u{g}}lu} S.,  {{\c{S}}a{\textcommabelow s}maz
  Mu{\textcommabelow s}} S.,   {Ek{\textcommabelow s}i} K.~Y.,  2020, \mn@doi
  [\mnras] {10.1093/mnras/staa1556}, \href
  {https://ui.adsabs.harvard.edu/abs/2020MNRAS.496.2183C} {496, 2183}

\bibitem[\protect\citeauthoryear{{{\c{S}}a{\textcommabelow s}maz
  Mu{\textcommabelow s}}, {{\c{C}}{\i}k{\i}nto{\u{g}}lu}, {Ayg{\"u}n},
  {Anda{\c{c}}}  \& {Ek{\textcommabelow s}i}}{{{\c{S}}a{\textcommabelow s}maz
  Mu{\textcommabelow s}} et~al.}{2019}]{Samaz2019}
{{\c{S}}a{\textcommabelow s}maz Mu{\textcommabelow s}} S.,
  {{\c{C}}{\i}k{\i}nto{\u{g}}lu} S.,  {Ayg{\"u}n} U.,  {Anda{\c{c}}} I.~C.,
  {Ek{\textcommabelow s}i} K.~Y.,  2019, \mn@doi [\apj]
  {10.3847/1538-4357/ab498c}, \href
  {https://ui.adsabs.harvard.edu/abs/2019ApJ...886....5S} {886, 5}

\makeatother
\end{thebibliography}

\end{document}